\documentclass[aps,english,floatfix,twocolumn,10pt]{revtex4-1}
\usepackage[T1]{fontenc}
\usepackage{amsmath}
\usepackage{textcomp}
\usepackage[latin1]{inputenc}
\usepackage{subfigure}
\usepackage{graphicx}
\usepackage{blindtext, rotating}
\usepackage{mathtools}
\usepackage{blindtext}
\usepackage{gensymb}
\usepackage{enumitem}
\usepackage{xcolor}
\usepackage{babel}
\usepackage{hyperref}
\usepackage{braket}
\usepackage{natbib}
\usepackage{subfigure}  
\makeatother
\begin{document}

\title{ Half-metallicity and two-dimensional hole gas at the $\text{BiFeO}_\text{3}$(001) surface }

\date{\today}
\author{Soumyasree Jena}
\email{jenasoumyasree@gmail.com}
\affiliation{Department of Physics and Astronomy, National Institute of Technology, Rourkela, Odisha, India, 769008}
\author{Sanjoy Datta} 
\email{dattas@nitrkl.ac.in}
\affiliation{Department of Physics and Astronomy, National Institute of Technology, Rourkela, Odisha, India, 769008}
\affiliation{Center for Nanomaterials, National Institute of Technology, Rourkela, Odisha, India, 769008}

\begin{abstract}
\noindent
The electronic structure and thermodynamic stability of tetragonal $\rm{BiFeO_3}$(001) surfaces have been 
investigated using density functional theory. In this work, four different structures having different lattice constants with 
two possible surface terminations have been studied. We have found that the surface electronic structure and the thermodynamic stability is quite 
sensitive with respect to the nature of the surface termination. The $\rm{FeO_2}$ terminated surface is found to be energetically 
more stable compared to $\rm{BiO}$ terminated surface in all the cases. Interestingly, we have found evidences of half-metallicity 
and spin-polarized two-dimensional hole gas (2DHG) at the one mono-layer thick surface in all the structures. 
The effect of the surface thickness have been systematically studied.
It has been demonstrated that the half-metallic 2DHG survives 
only in one of the structures for all the thicknesses, incidentally, which is the most thermodynamically stable structure.\\

\textbf {Keywords:} Tetragonal $\rm{BiFeO_3}$(001), Half-metallicity, 2DHG 
\end{abstract}
\maketitle
\section{Introduction}
The study of two-dimensional systems like oxide interfaces and the surface analysis of perovskite systems have 
received wide attention since the discovery of the two dimensional electron gas (2DEG) in the  $\rm{SrTiO_3}/\rm{LaAlO_3}$ 
hetero-interface\cite{2deg} 
and its counterpart i.e., two dimensional hole gas (2DHG) in the  $\rm{SrTiO_3}$/$\rm{LaAlO_3}$/$\rm{SrTiO_3}$\cite{2dhg}. 
In addition, the 2DHG is also found in 
$\rm{SrTiO_3}$ interface\cite{sto-interface}.  These findings open up 
new windows of opportunities from the point of view of designing novel functional devices.   
In 2006, R. Pentcheva and W. E. Pickett\cite{pickett-hole-polaron}
reported the presence of magnetic holes in p-type interface of $\rm{SrTiO_3}$/$\rm{LaAlO_3}$. Interestingly, the half-metallic 2DHG is found in the 
$\rm{SrTiO_3/KTaO_3}$ interface \cite{sto/kto-2dhg}.

Following these findings, the search for 2DEG and 2DHG has been extended to other systems, for example, 
existence of 2DEG has been experimentally found in the surface of $\rm{KTaO_3}$ \cite{kto-2deg} when the electrons are confined 
along the $(001)$ direction. On the other hand, based on density functional theory based first principle calculations, 
existence of 2DEG has been predicted in the slab structures of the polar perovskites, such as $\rm{BaBiO_3}$ \cite{BaBio3-2deg} and
in $\rm{SrTiO_3}(111)$ \cite{sto-111}\cite{sto-111a}. Apart from these, there have been several other studies on other 
polar perovskites, for example the stability and structural properties of $\rm{BaTiO_3}(110)$ slabs have been studied 
in Ref.~\cite{bto-110}, polar surface structures of $\rm{PbTiO_3}$, $\rm{SrZrO_3}$ and $\rm{PbZrO}_3(111)$ were studied 
in Ref.~\cite{polar-compounds}, while the surface properties of rhombohedral $\rm{BiFeO_3}$ has been investigated 
in Ref.~\cite{rbfo-0001}. In addition, the rhombohedral $\rm{BiFeO_3}$ thin film is studied experimentally in Ref.~\cite{rbfo-expt}.
However, 2DHG has only been found in a limited set of systems as compared to the 2DEG. 
Recently, it has been reported that 2DHG co-exists with 2DEG in the strained polar perovskite thin film of 
$\rm{KTaO_3}$\cite{kto-tf}. 

Motivated by these studies, in this work, we have investigated the slab structures of tetragonal $\rm{BiFeO_3}(001)$ (TBFO), which is 
also a polar perovskite. Recently, bulk TBFO\cite{bulk-paper} has been found to host wide range of electronic phases 
in the presence of ferromagnetic ordering. In the past, TBFO has been fabricated having a wide range of the $c/a$ 
ratios\cite{bulk-paper}. Interestingly, it has been found that certain bulk structures of TBFO can become half-metallic \cite{groot} in the presence of 
ferromagnetic ordering of the spins. Furthermore, in the half-metallic phase of the bulk  system the charge carriers were found to be of electron type.
\begin{figure}[htbp!]
\centering
	\includegraphics[width=0.45\textwidth,height=0.40\textwidth]{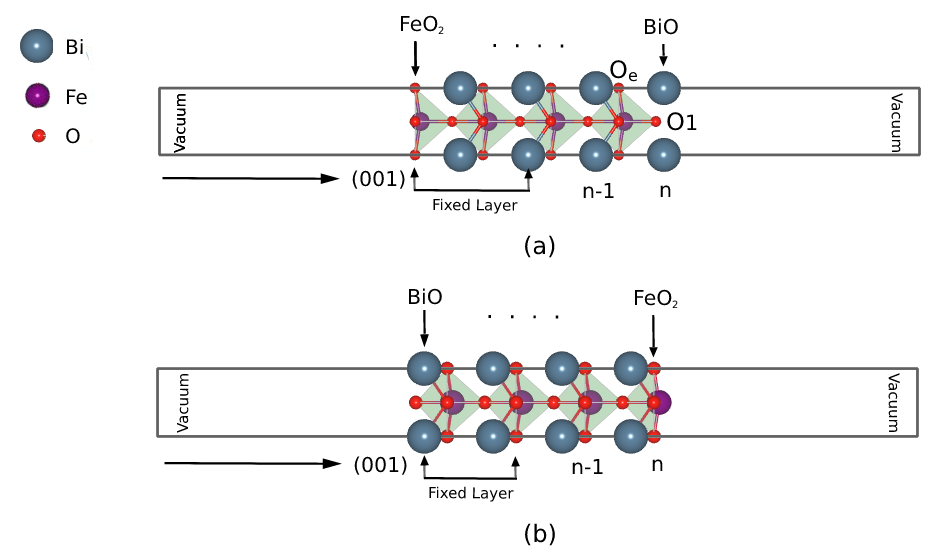}
\vspace{-0.4cm}
\caption{(a) $\rm{BiO}$ terminated, and (b) $\rm{FeO_2}$ terminated slab with 2 mono-layered thick surfaces respectively.}
\label{Fig:cell-structure}
\end{figure}

\begin{figure}[htbp!]
\centering
	\includegraphics[width=0.45\textwidth,height=0.18\textwidth]{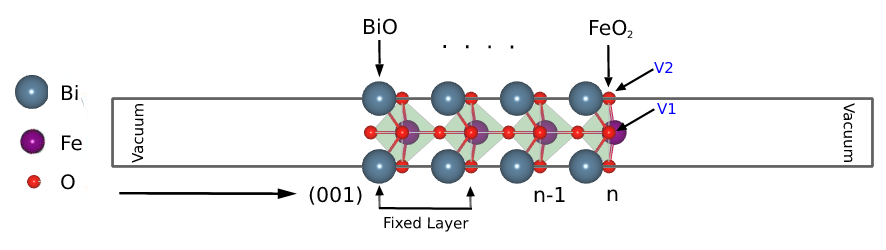}\hspace{2.01cm}
\vspace{-0.4cm}
\caption{Oxygen vacancy on the 2 mono-layered thick $\rm{FeO_2}$ terminated surface with at site-1(V1) ,
and at site-2(V2) respectively.}
\label{Fig:ov}
\end{figure}
Naturally the question arises whether the half-metallic phase also survives on the surface of a TBFO
slab structure, and, if it survives, are the charge carriers still electron type or the reduced dimensionality brings 
in some drastic changes in its nature. To address these questions, we have systematically studied the surface of the 
slab geometries with increasing thickness (1, 2 and 3 monolayers) for 
four different structures (I, II, III, and IV) of TBFO having different lattice parameters \cite{bulk-paper}. 
Furthermore, the TBFO-slabs have been examined with positive ($\rm{BiO}$) and negative ($\rm{FeO_2}$) surface terminations.
The electronic properties of the surface have been found to be drastically influenced by the nature of the termination layer. All the structures
have been found to host metallic surface states when the surface is positively terminated. However, when the surface is negatively terminated, 
the nature of the surface states have been found to be sensitive to the thickness of the slab as well as on the lattice parameters. 
Out of all the structures, the surface states of the structure II have been found to possesses the highest robustness with respect to the change 
in the slab 
thickness. Interestingly, the surface of this structure has been found to be \textit{half-metallic} similar to its bulk counterpart. 
Surprisingly, however, the charge carriers become hole type in the slab, in contrast to the bulk structure, indicating the existence 
of spin-polarized 2DHG at the surface of TBFO slab. It is important to note that, 
half-metallic ferromagnetism (HMFM) has also been reported in TBFO based hetero-structure systems 
\cite{lsmo/bfo}\cite{lsmo/bfo-hm}\cite{2d-pbo/bfo-hm}. Also, existence of 2DEG has been predicted in 
$\rm{TBFO}/\rm{SrTiO_3}$ heterostructure\cite{bfo-sto}. 
To the best of our knowledge, possibility of 
spin-polarized 2DHG in pure TBFO slab is being reported for the first time in this work. 
\begin{figure*}[htbp!]
\centering
	\includegraphics[width=0.26\textwidth,height=0.26\textwidth]{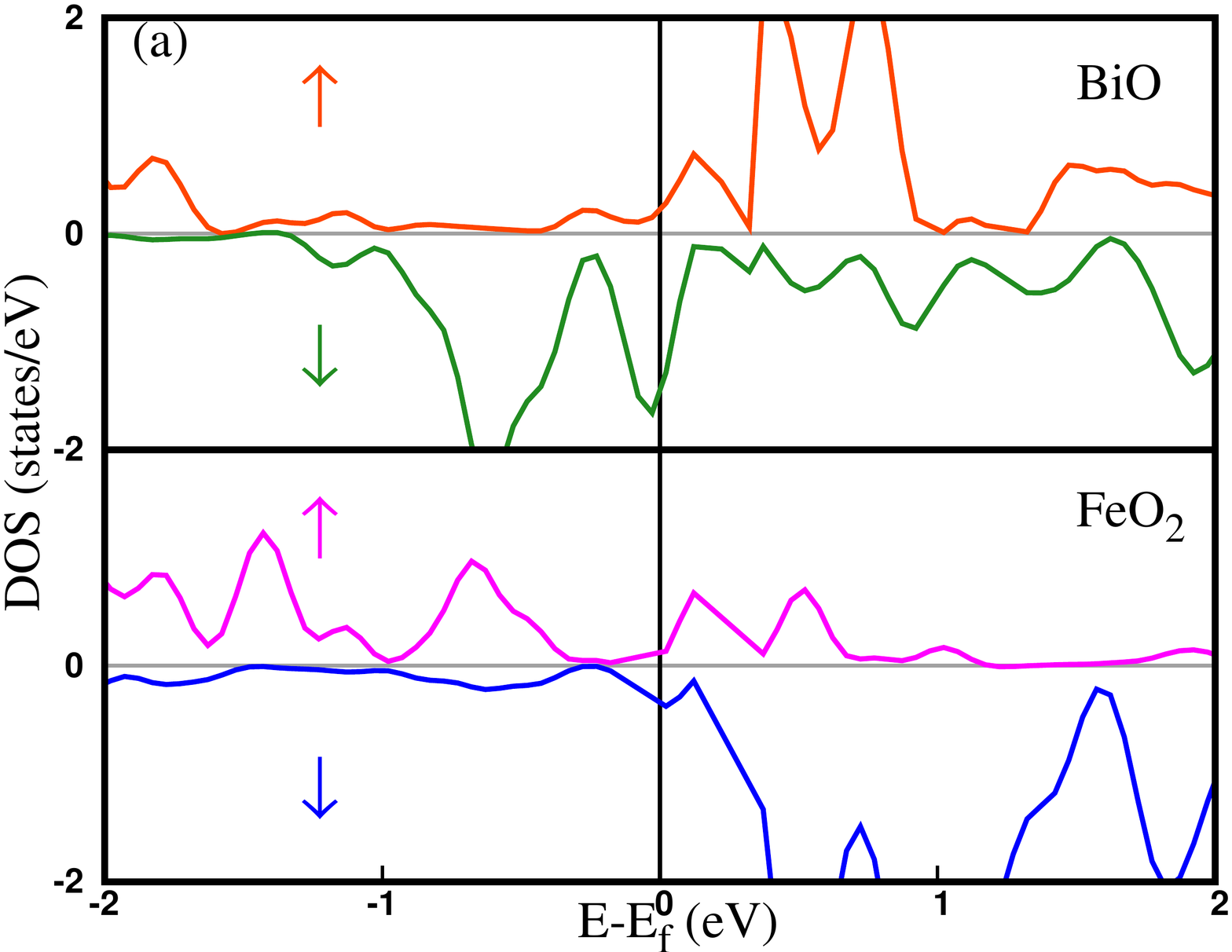}\hspace{-0.5cm}
	\includegraphics[width=0.26\textwidth,height=0.26\textwidth]{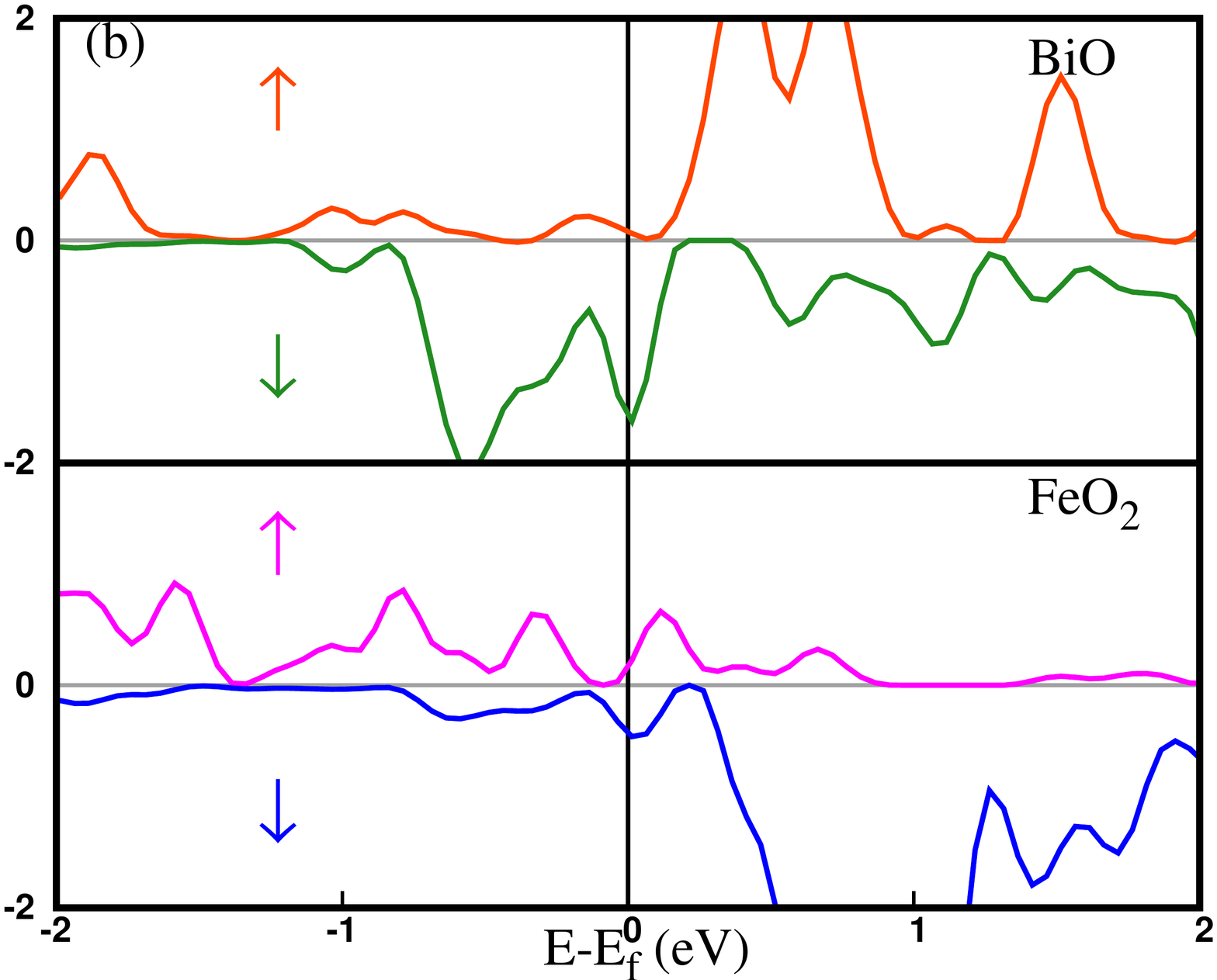}\hspace{-0.5cm}
	\includegraphics[width=0.26\textwidth,height=0.26\textwidth]{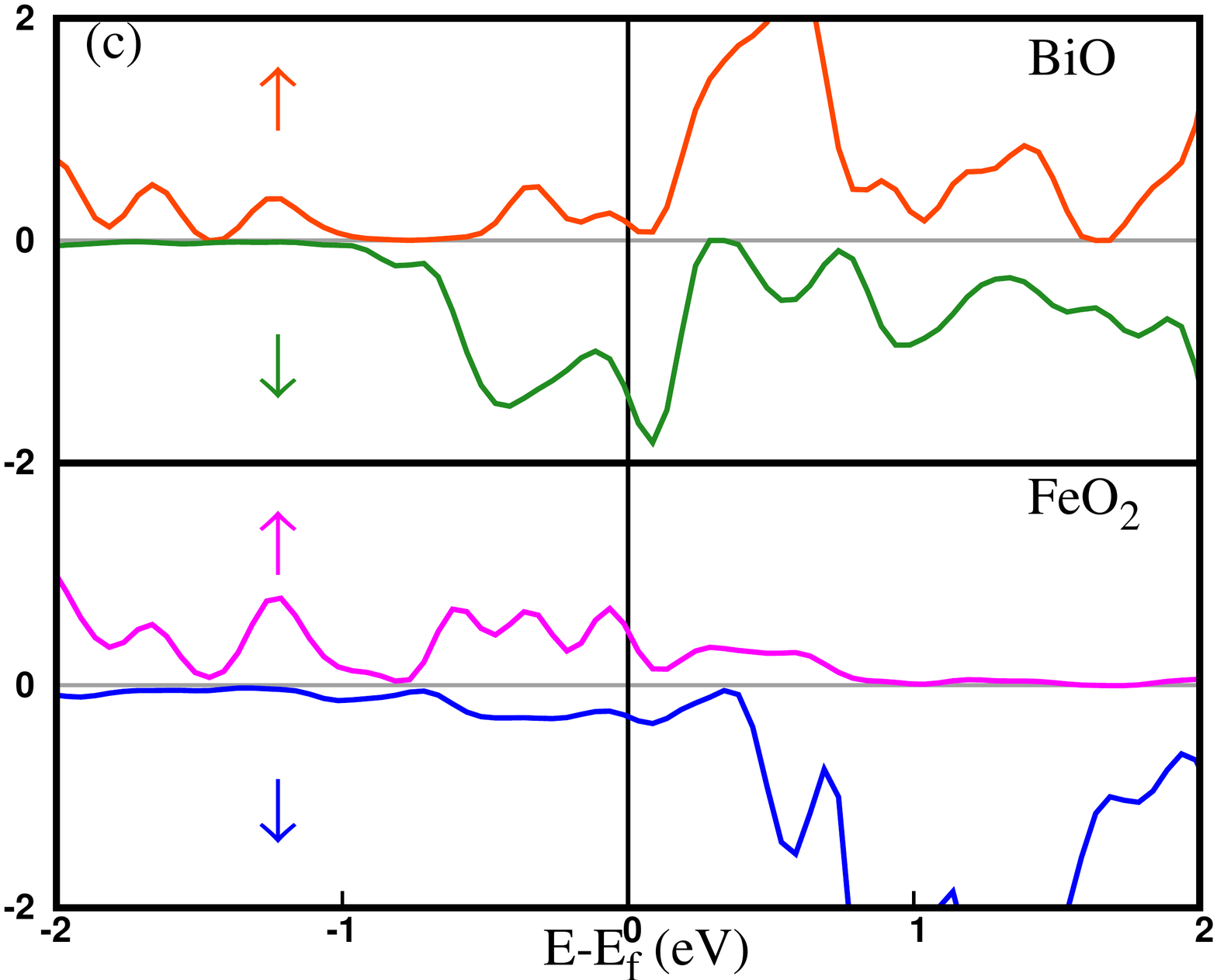}\hspace{-0.5cm}
	\includegraphics[width=0.26\textwidth,height=0.26\textwidth]{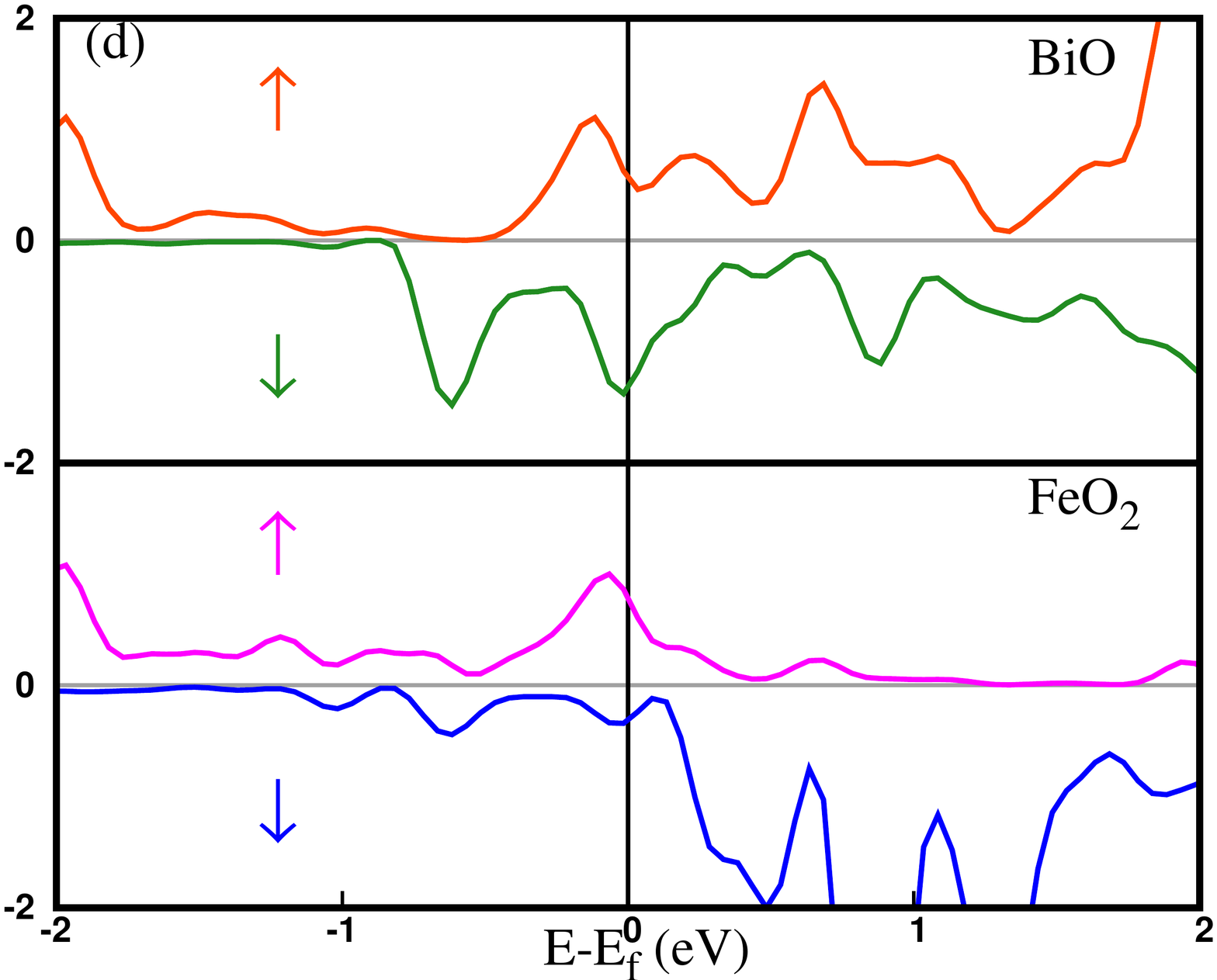}
\vspace{-0.4cm}
\caption{Layered density of states (DOS) of 1ml-thickness slab with $\rm{BiO}$ termination (a) Structure-I, 
(b) Structure-II , (c) Structure-III, (d) Structure-IV respectively.}
\label{Fig:dos-bio-1l}
\end{figure*}

\begin{figure*}[htbp!]
\centering
	\includegraphics[width=0.26\textwidth,height=0.26\textwidth]{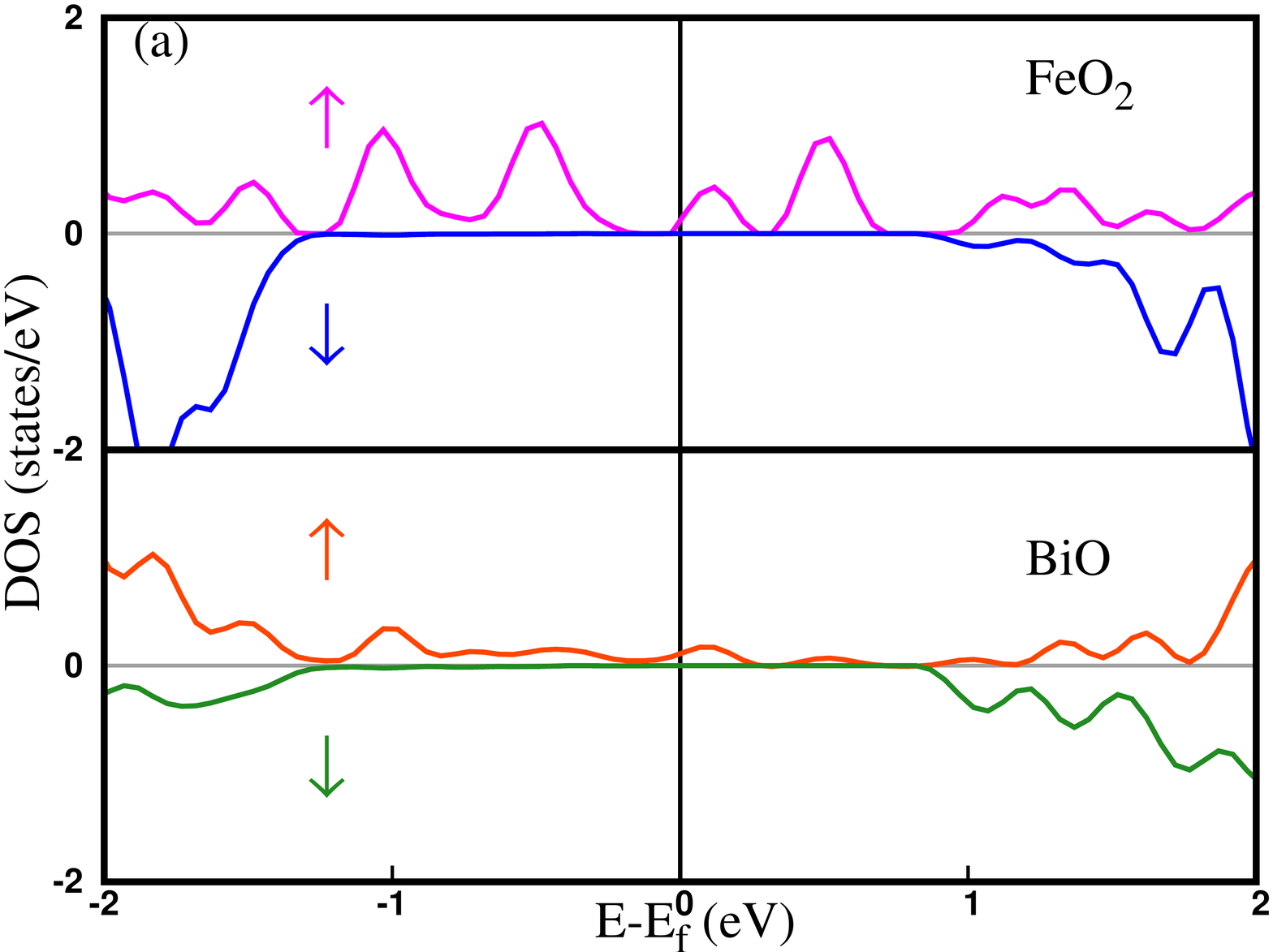}\hspace{-0.5cm}
	\includegraphics[width=0.26\textwidth,height=0.26\textwidth]{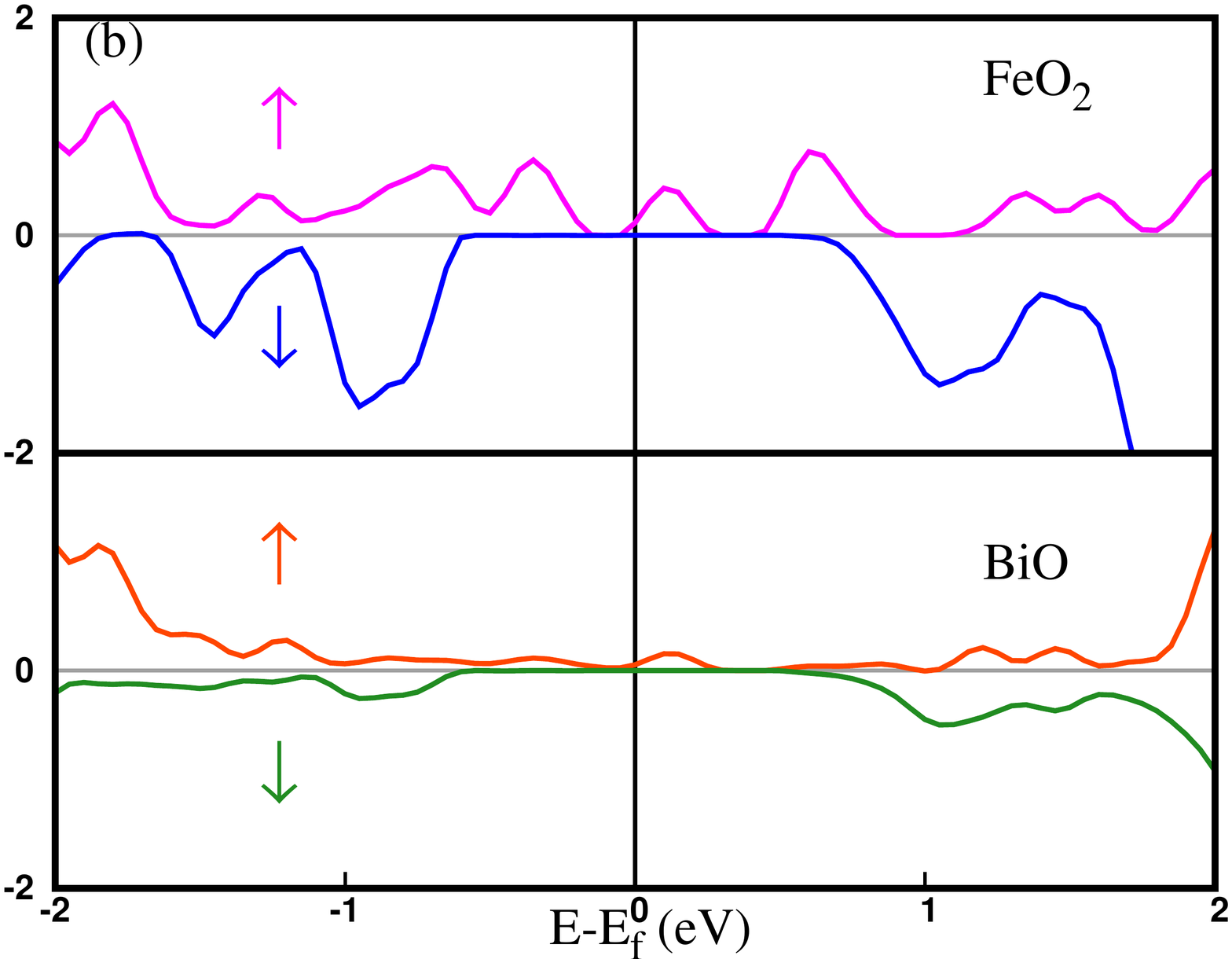}\hspace{-0.5cm}
	\includegraphics[width=0.26\textwidth,height=0.26\textwidth]{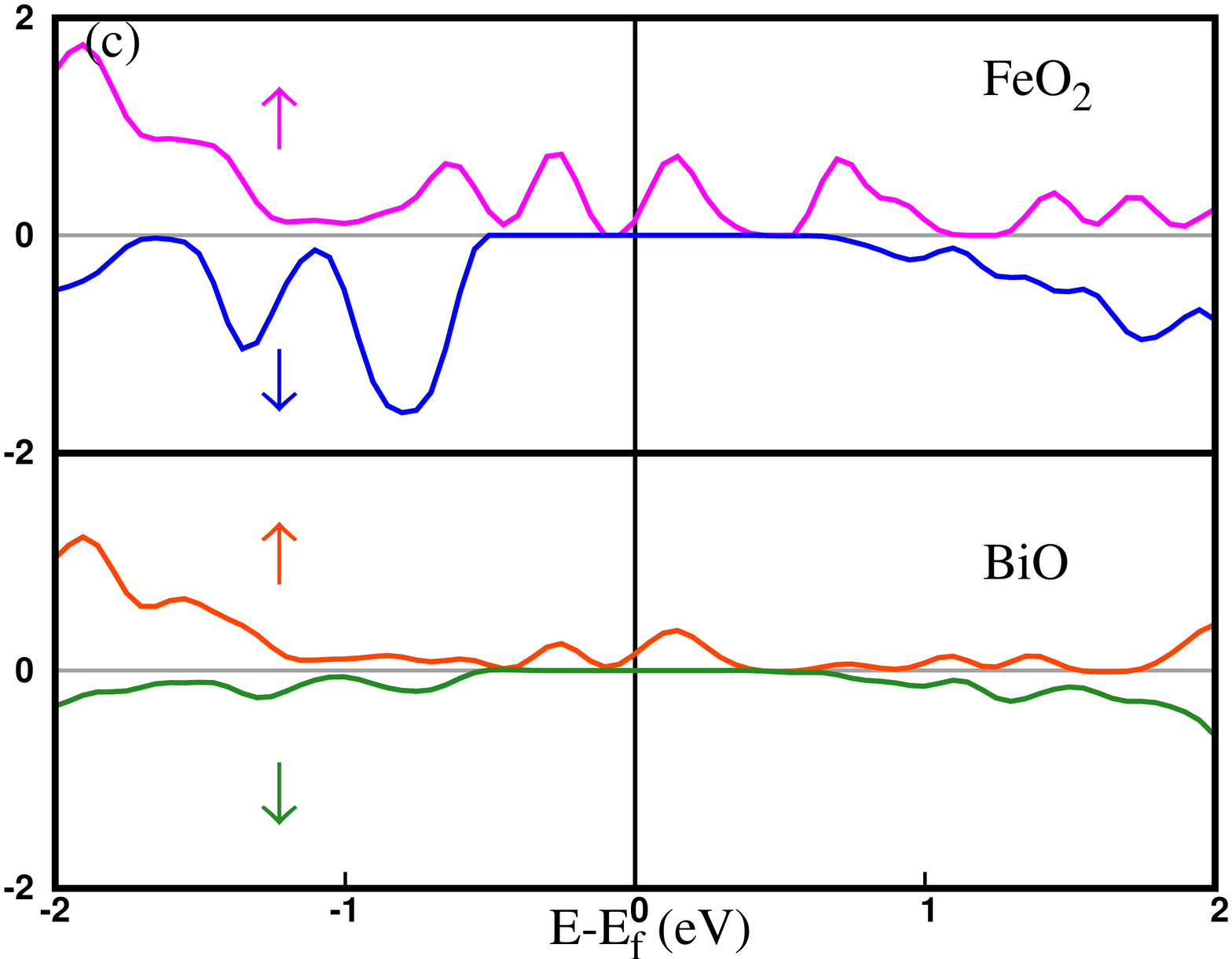}\hspace{-0.5cm}
	\includegraphics[width=0.26\textwidth,height=0.26\textwidth]{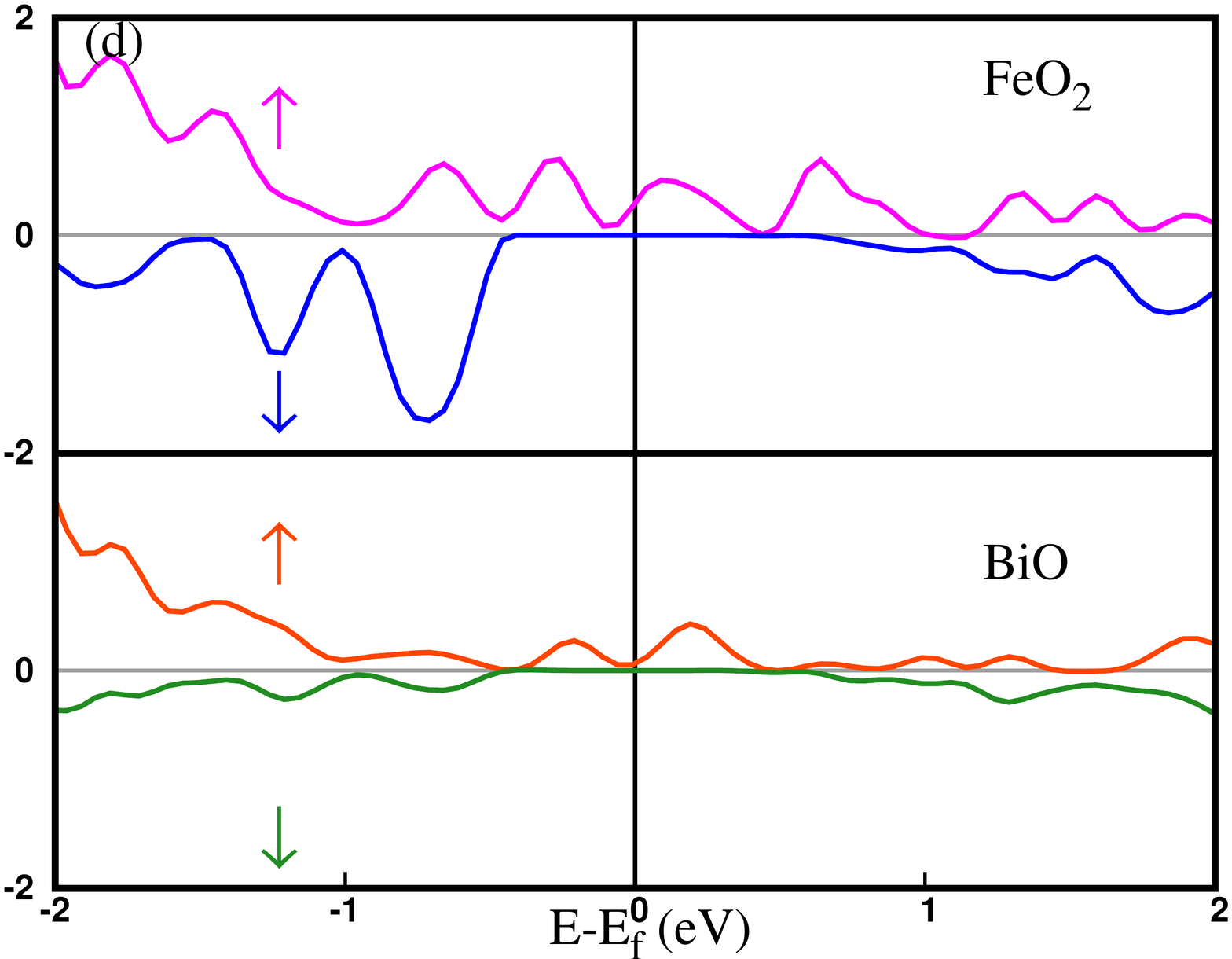}\hspace{-0.5cm}
\vspace{-0.4cm}
\caption{Layered density of states (DOS) of 1ml slab-thicknesses with $\rm{FeO_2}$ termination for (a) structure-I, (b) structure-II, 
(c) structure-III and (d) structure-IV respectively.} 
\label{Fig:1l-dos-feo2}
\end{figure*}
The paper is organized as follows: Computational and structural details are provided in Sec.~\ref{computational deatails} and Sec.\ref{structural deatails}
followed by the thermodynamic stability analysis of the slab structures in Sec.~\ref{thermodynamic stability}. The detail 
study of the electronic and magnetic properties are given in the Sec.~\ref{results}. The results of stoichiometric 
slabs have been discussed in Sec.~\ref{stoichiometric slabs}. In Sec.~\ref{oxygen-vacancy}~, we have demonstrated
the effect of oxygen-vacancies on the structure II. For the stoichiometric slabs, the $\rm{BiO}$ terminated 
systems have been studied in Sec.\ref{bio-termination}, while the results and discussion of $\rm{FeO_2}$ terminated systems are presented in 
Sec.\ref{feo2-termination}. Finally, we have concluded our work in the Sec.~\ref{conclusions}.
\section{Methods and Calculations}
\subsection{Computational details}\label{computational deatails}
For all the computational calculations, the spin-polarized DFT based first-principle calculations with plane-wave basis set is performed,
which is implemented in the QUANTUM ESPRESSO package \cite{qe}. The projected augmented wave (PAW) \cite{paw}
method obtained pseudopotentials are used in which 15 valence electrons 
$(5d^{10}6s^{2}6p^{3})$ for Bismuth (Bi), 16 valence electrons $(3s^{2}3p^{6}3d^{6}4s^{2})$ for Iron(Fe) and 6 valence electrons
$(2s^{2}2p^{4})$ for Oxygen(O) have been used. The Perdew-Burke-Ernzerhof (PBE) \cite{pbe} type GGA+U i.e., generalized gradient approximations 
is employed with a Hubbard U value 4.5 eV\cite{bulk-paper}, which is applied to Fe-3d orbital.
A vacuum of 15\AA{} has been used to avoid any spurious dipole-dipole interaction.
The $6\times6\times1$ mp-grid \cite{mp} is used for the self-consistent calculation. 
The K-points are chosen by following the 
criterion of  0.01 ~meV/atom. The electronic self consistent calculations are done for a total energy 
convergence of less than $10^{-7}$ eV. An ionic relaxation is performed to optimize the inter-atomic
forces less than $10^{-3}$ Ry/bohr. The Methfessel-Paxton \cite{Methfessel-paxton} type smearing is used in all the calculations.
The kinetic energy cut off i.e., $\rm{E_{cut}}$ has been set to be 80 Ry.
For all the calculations, spin-orbit coupling is ignored.
\subsection{Structural details}\label{structural deatails}
We have systematically studied the TBFO (001) slab structures in the presence of the FM ordering.
For this study we have used asymmetrical slab geometries that is the top and the bottom surfaces are 
oppositely charged. 
The unit-cell TBFO carries the $\rm{C_{4v}}$ point-group symmetry. 
Typically these slab structures are polar in nature, and in this case it comprises of two complementary slices, i.e., 
a positively charged $\rm{(BiO)}^{+}$ and a negatively charged $\rm{(FeO_2)}^{-}$ layer in the $[001]$ direction.
In Fig.~\ref{Fig:cell-structure}. the slab structure having
two surface mono-layers with two different possible terminations have been presented.  

The results presented in this article are for structures obtained after performing ionic relaxation. In the process of 
ionic relaxation we have kept half of the layers from top to be free for relaxation and rest of the layers 
to be fixed i.e., clamped to the bulk co-ordinates since our primary focus is to understand the 
electronic properties of the top surface layers.

In Fig.~\ref{Fig:cell-structure}, 
the snapshot of the two mono-layered (2ml) slabs are given where the surface layers are enumerated. Here, \textit{`n'} and \textit{`n-1'} stand for the top 
and bottom surface layers respectively. 
For a systematic study of the 
thickness dependence of the electronic properties we have used structures having 1 (1ml), 2 (2ml) and 3 monolayers (3ml). 
Furthermore, we have used four different structures of TBFO having lattice constant (\textit{`a'}) 
3.670\cite{dr}, 3.770, 3.880 and 3.935 \AA{}\cite{wang} \cite{gp}. These structures are labeled as 
structure-I, structure-II, structure-III and structure-IV respectively.
\begin{figure*}[htbp!]
\centering
	\includegraphics[width=0.26\textwidth,height=0.26\textwidth]{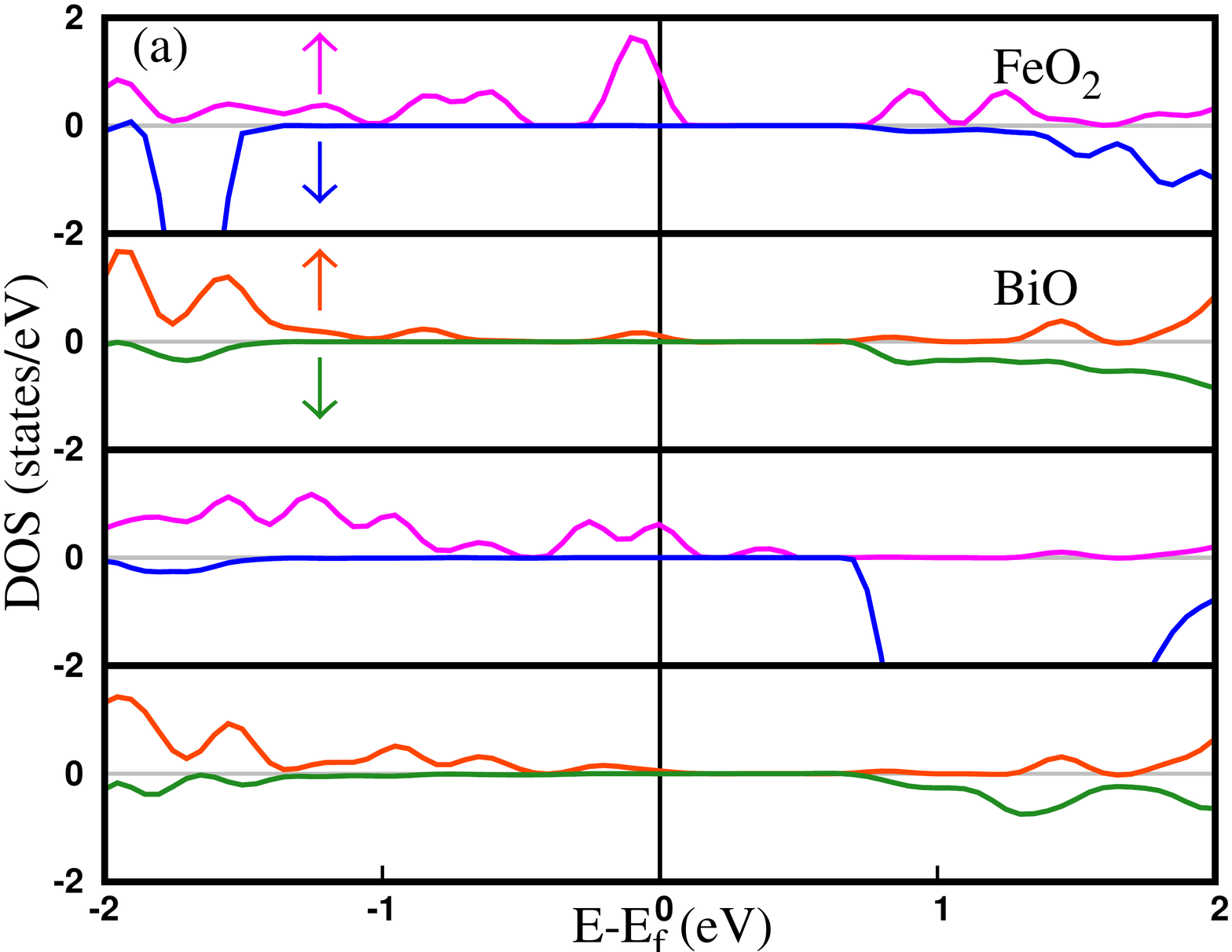}\hspace{-0.5cm}
	\includegraphics[width=0.26\textwidth,height=0.26\textwidth]{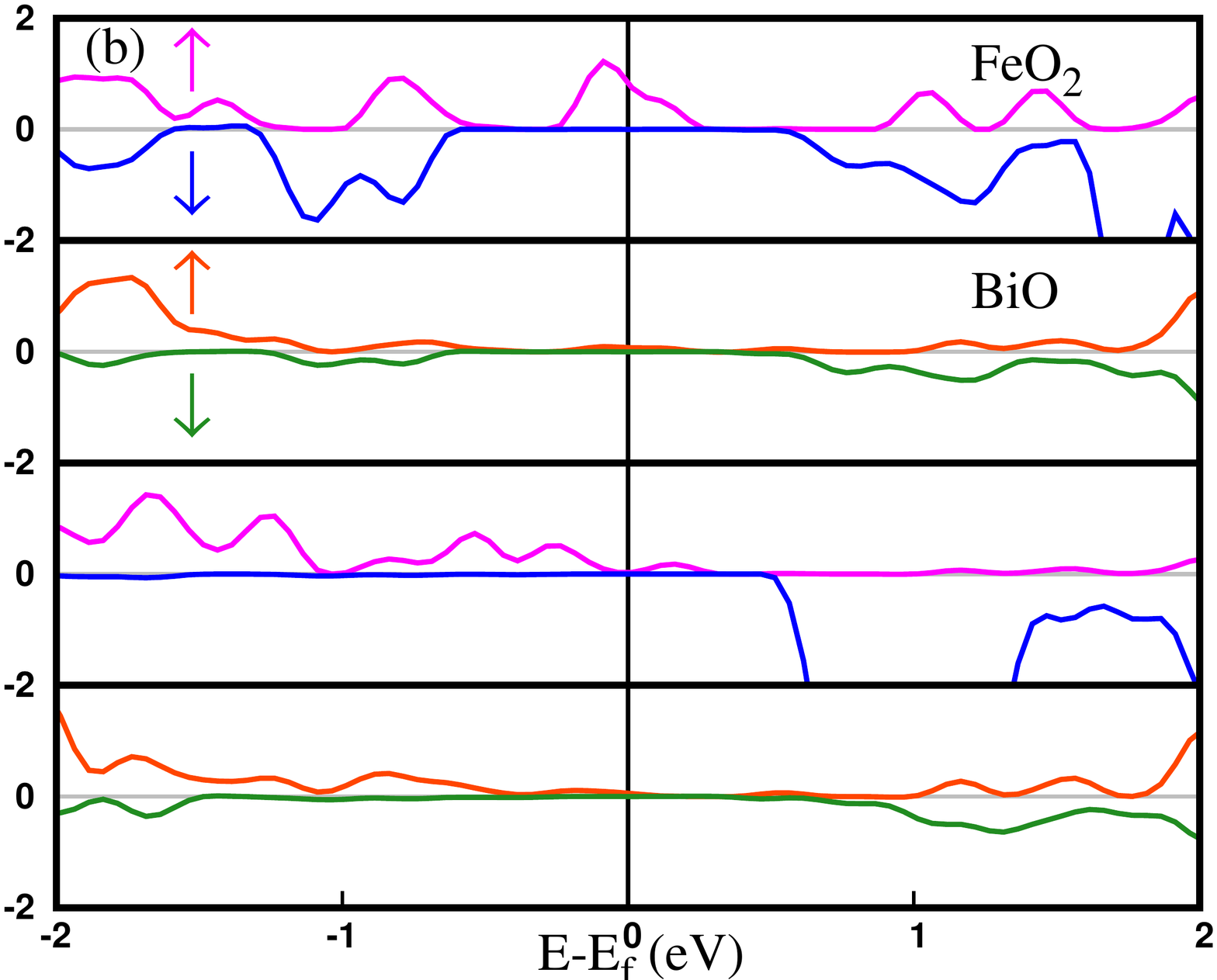}\hspace{-0.5cm}
	\includegraphics[width=0.26\textwidth,height=0.26\textwidth]{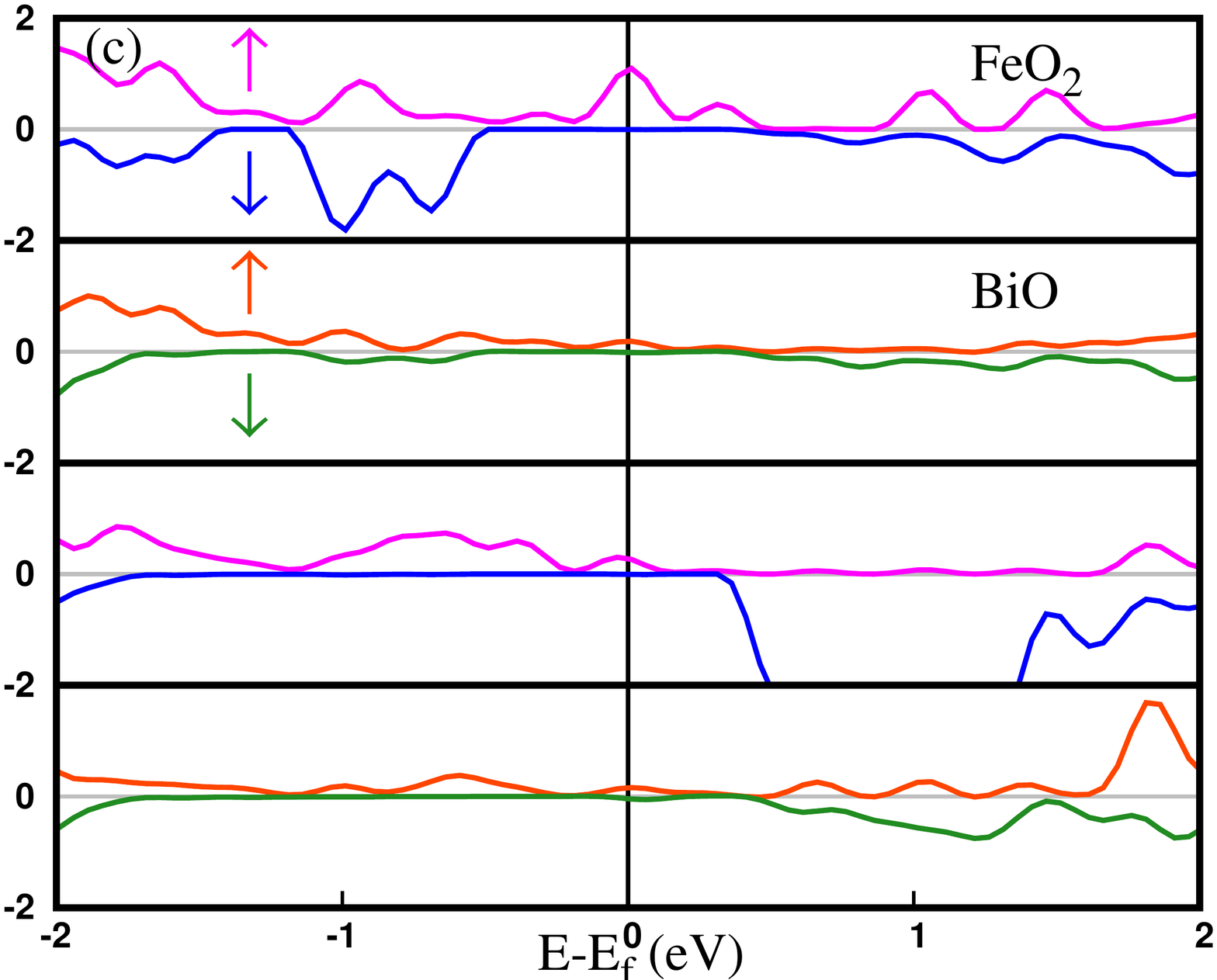}\hspace{-0.5cm}
	\includegraphics[width=0.26\textwidth,height=0.26\textwidth]{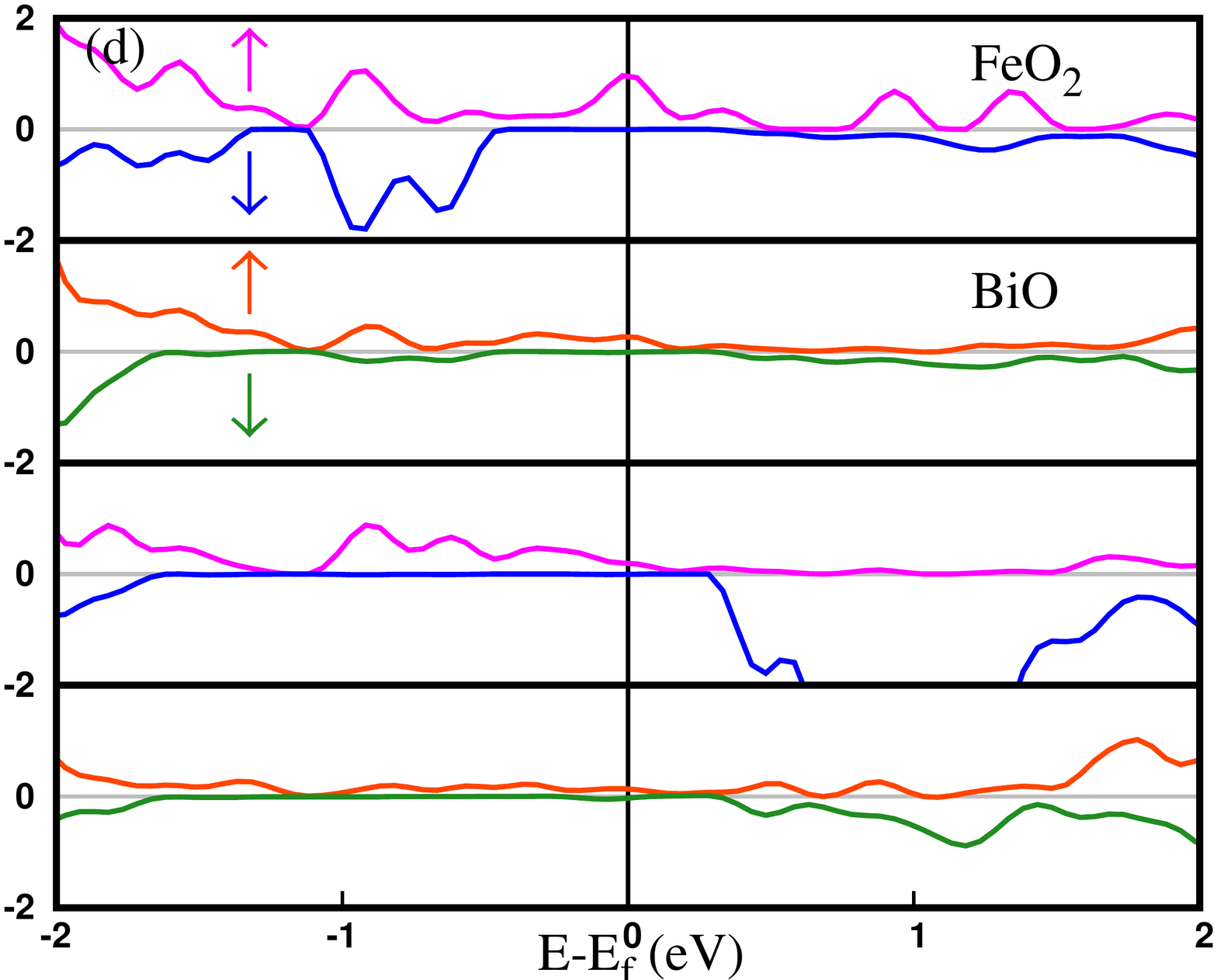}\hspace{-0.5cm}
\vspace{-0.4cm}
\caption{Layered density of states (DOS) of 2ml slab-thicknesses with $\rm{FeO_2}$ termination (a) structure-I, (b) structure-II, 
(c) structure-III and (d) structure-IV respectively.} 
\label{Fig:2l-dos-feo2}
\end{figure*}
 \begin{figure*}[htbp!]
\centering
	\includegraphics[width=0.26\textwidth,height=0.26\textwidth]{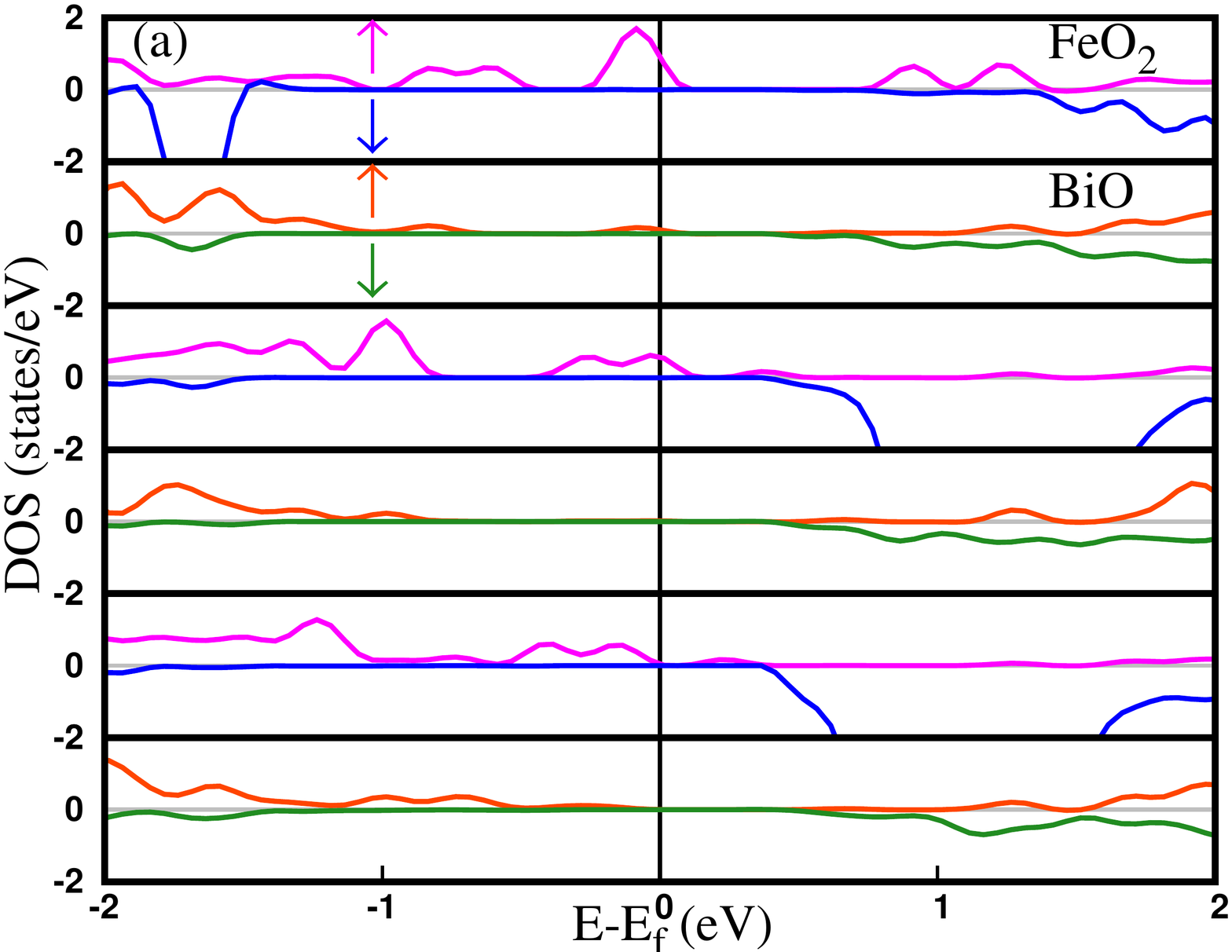}\hspace{-0.5cm}
	\includegraphics[width=0.26\textwidth,height=0.26\textwidth]{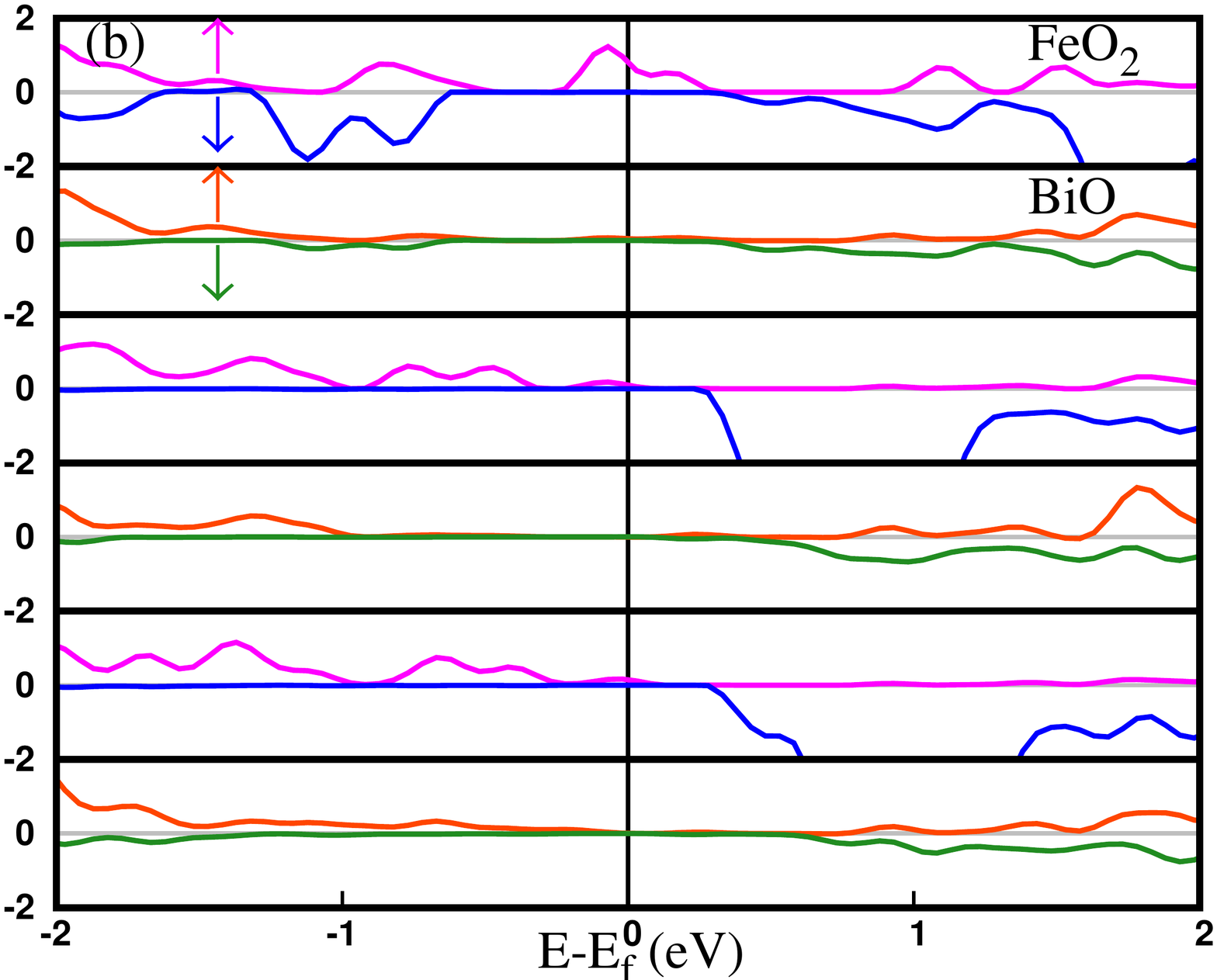}\hspace{-0.5cm}
	\includegraphics[width=0.26\textwidth,height=0.26\textwidth]{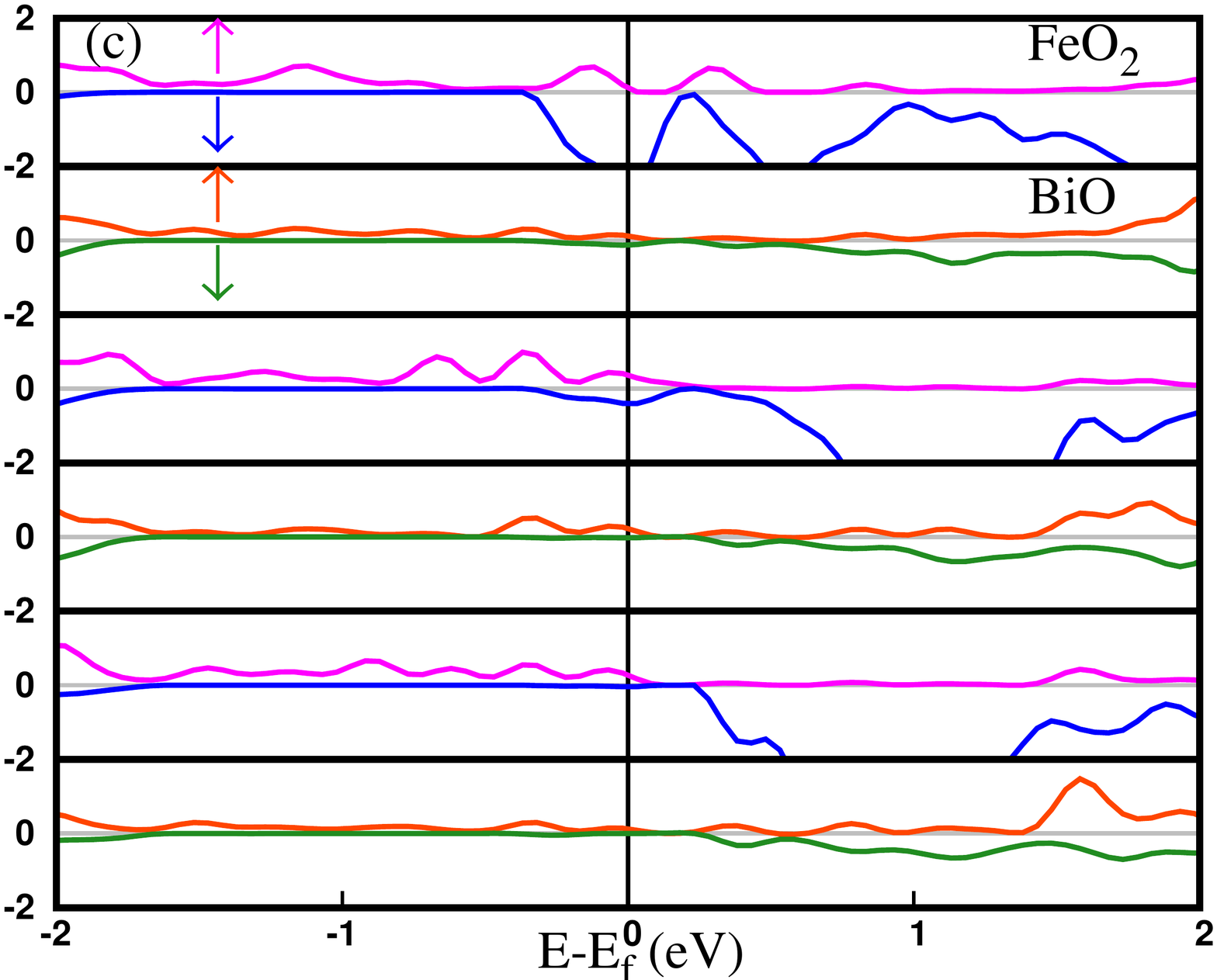}\hspace{-0.5cm}
	\includegraphics[width=0.26\textwidth,height=0.26\textwidth]{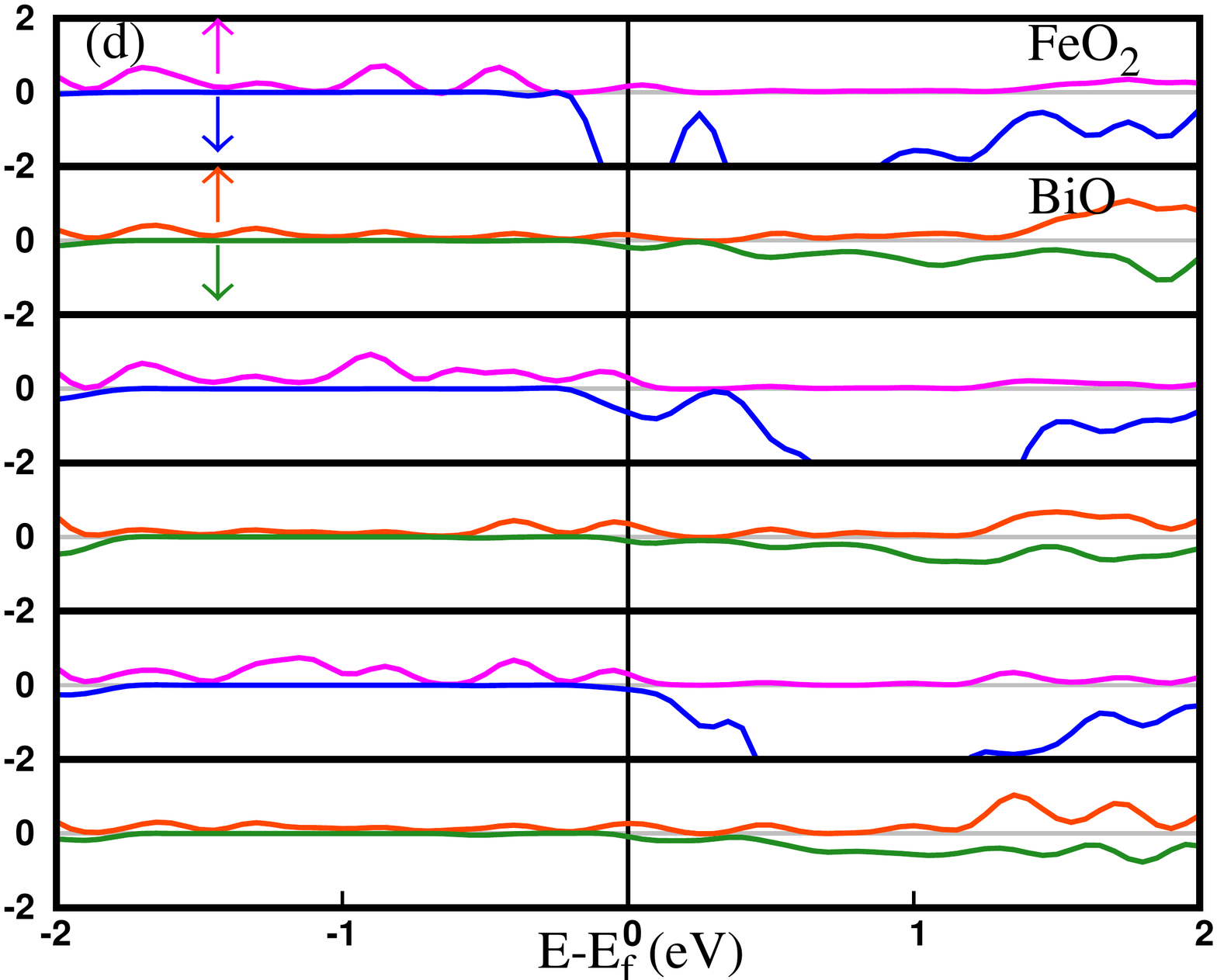}\hspace{-0.5cm}
\vspace{-0.4cm}
\caption{Layered density of states (DOS) of 3ml  slab-thicknesses with $\rm{FeO_2}$ termination  (a) structure-I, (b) structure-II, 
(c) structure-III and (d) structure-IV respectively.} 
\label{Fig:3l-dos-feo2}
\end{figure*}
In our study, structure-II has been found to be most stable and it also hosts 
the spin-polarized 2DHG at the surface.

The effect of oxygen vacancy on the surface 
states has been studied, particularly for structure-II. There are two inequivalent sites at which the oxygen vacancies may appear at the top surface.   
The two different sites for oxygen vacancies, which are named as V1 and V2, have been presented in Fig.~\ref{Fig:ov}.
Our calculations have been carried out for the $\rm{FeO_2}$ terminated slab for all the thicknesses in order 
to understand the persistence of the 2DHG at the surface.
\section{Thermodynamic stability}\label{thermodynamic stability}

Before presenting the results of the electronic properties, in this section we discuss the thermodynamic stability of the TBFO slabs.
The stability of the surface with respect to the bulk can be studied by estimating the surface energy. It is the sum of the 
cleavage energy (in unit $J/m^{2}$) of the unrelaxed surface and the relaxation energy (in unit $J/m^{2}$). The surface energy $\rm{E}_s$ 
is defined as, 
\begin{equation}
\rm{E_s = E_{cl} + E_{rel}},
\label{Eq:surface-energy}
\end{equation}
where $\rm{E}_{cl}$ and $\rm{E}_{rel}$ are the cleavage and relaxation energies respectively. 
The cleavage energy can be defined as the amount of energy required to break a crystal into 
complementary slices mentioned above. Stoichiometrically, in [001] direction, the TBFO possesses two complementary unrelaxed 
surfaces i.e., $\rm{FeO_2}$ and $\rm{BiO}$, which can be obtained by cleaving the bulk crystal.
The cleavage energy for the two possible slab structures have been defined in 
Eqs.~\ref{Eq:CleavageEnery-1} and \ref{Eq:CleavageEnery-2} as,
\begin{equation}
E^{FeO_2/BiO}_{cl}=\frac{1}{2S}(E^{FeO_2}_{slab}+E^{BiO}_{slab}-nE_{bulk}),
\label{Eq:CleavageEnery-1}
\end{equation}
\begin{equation}
E^{BiO/FeO_2}_{cl}=\frac{1}{2S}(E^{BiO}_{slab}+E^{FeO_2}_{slab}-nE_{bulk}).
\label{Eq:CleavageEnery-2}
\end{equation}
In the above equations, $E_{bulk}$ represents the total energy possessed by the bulk TBFO, \textit{`n'} denotes the total number of BFO 
monolayers and \textit{`S'} represents the surface area. It is clear that for the asymmetrical slab structures, the 
cleavage energies are same for both kind of 
surface termination. In Table~\ref{Table:cleavage-energies}, we have presented the cleavage energies for both kind of surface 
terminations and for different thickness of the TBFO slab.
\begin{table}[htbp!]
\caption{Estimation of the cleavage energies(in unit $\rm{J/m^2}$) for the two types of surface terminations.}
\begin{center}
\begin{tabular}{c c c c c c c }
\hline
Structures  & I  & II & III & IV   \\
\hline
1ml          & 2.15 & 2.61  & 2.99 & 3.12  \\
2ml          & 1.82 & 2.64  & 2.71 & 2.77  \\
3ml         & 1.85 & 2.63  & 2.72 & 2.83   \\
\hline
\label{Table:cleavage-energies}
\end{tabular}
\end{center}
\end{table} 

The relaxation energy i.e., $\rm{E_{rel}}$ can be defined as, 
\begin{equation}
 \rm{E_{rel} = \frac{1}{2S}[E^{rel}_{slab}(A) - E^{unrel}_{slab}(A)]},
\end{equation}
where, $\rm{E^{rel}_{slab}(A)}$ stands for the relaxed energy of the slab termination $\rm{A}$ and $\rm{E^{unrel}_{slab}(A)}$ stands for 
the unrelaxed energy for the slab with the same termination layer. The estimation of the relaxation energies for the two types of 
termination layer are presented in Tables~\ref{Table:bio-relaxation-energy} and \ref{Table:feo2-relaxation-energy} respectively. 
\begin{table}[htbp!]
\caption{Estimation of the relaxation energies(in unit $\rm{J/m^2}$) for the $\rm{BiO}$ surface termination.}
\begin{center}
\begin{tabular}{c c c c c c c }
\hline
Structures  & I  & II & III & IV   \\
\hline
1ml          & -0.81  & -0.42  & 0.05 & 0.11 \\
2ml          & -1.86  & -0.94  & 0.39 & 0.41  \\
3ml         & -2.86 & -1.57  & 0.41 & 0.36 \\
\hline
\label{Table:bio-relaxation-energy}
\end{tabular}
\end{center}
\end{table}
\begin{table}[htbp!]
\caption{Estimation of the relaxation energies(in unit $\rm{J/m^2}$) for the $\rm{FeO_2}$ surface termination.}
\begin{center}
\begin{tabular}{c c c c c c c }
\hline
Structures  & I  & II & III & IV   \\
\hline
1ml          & -1.53 & -0.59  & -0.19 & -0.23 \\
2ml          & -2.33 & -1.16  & 0.14 & 0.21 \\
3ml         & -3.41 & -1.78  & 0.48 & 0.62 \\
\hline
\label{Table:feo2-relaxation-energy}
\end{tabular}
\end{center}
\end{table}
Finally, in Tables~\ref{Table:bio-surface-energy} and \ref{Table:feo2-surface-energy} our estimation of the surface energies of all the 
structures for the $\rm{BiO}$ and $\rm{FeO_2}$ surface termination have been presented .
\begin{table}[htbp!]
\caption{Estimation of the surface energies(in unit $\rm{J/m^2}$) for $\rm{BiO}$ surface termination.}
\begin{center}
\begin{tabular}{c c c c c c c }
\hline
Structures  & I  & II & III & IV   \\
\hline
1ml          & 1.34  & 2.19 & 3.04 & 3.23 \\
2ml          & -0.04 & 1.70 & 3.10 & 3.18 \\
3ml         & -1.01 & 1.06 & 3.13 & 3.19 \\
\hline
\label{Table:bio-surface-energy}
\end{tabular}
\end{center}
\end{table}
\begin{table}[htbp!]
\caption{Estimation of the surface energies(in unit $\rm{J/m^2}$) for $\rm{FeO_2}$ surface termination.}
\begin{center}
\begin{tabular}{c c c c c c c }
\hline
Structures  & I  & II & III & IV   \\
\hline
1ml          & 0.62 & 2.02  & 2.80 & 2.89 \\
2ml          & -0.51 & 1.48 & 2.85 & 2.98 \\
3ml         & -1.56 & 0.85 & 3.20 & 3.46 \\
\hline
\label{Table:feo2-surface-energy}
\end{tabular}
\end{center}
\end{table}

From the data of Tables \ref{Table:bio-surface-energy} and \ref{Table:feo2-surface-energy}, we can observe that the surface energy of
structure-I turns out to be negative as the thickness of the surface is increased beyond 1ml. This indicates the possibility
of instability in this structure \cite{alumina-negative-energy}. Apart from these particular cases, the surface energies 
have been found to be positive in all other cases. It is important to note that from the point of view of surface energy 1ml thick 
slab of structure-I with $\rm{FeO_2}$ termination is the most favorable structure. However, as we move beyond 1ml thickness, 
the surface energy of the structure-II becomes the lowest among all the structures. Furthermore, it can also be observed that the 
surface energies are invariably lower when the surface is $\rm{FeO_2}$ terminated. Interestingly, it is this $\rm{FeO_2}$ terminated 
structure-II in which we have found the most robust signature of half-metallicity and spin-polarized 2DHG at the surface. Half-metallicity and 
spin-polarized hole carriers have been found in structure III and IV as well. However, they do not survive in 3ml thick slab structure. 
\section{Results and Discussions}\label{results}
\subsection{Stoichiometric slabs}\label{stoichiometric slabs}
\subsubsection{$\rm{BiO}~termination$}\label{bio-termination}
In the thermodynamic analysis of Sec.~\ref{thermodynamic stability} it is found that the $\rm{FeO_2}$ terminated slabs
are more stable compared to the $\rm{BiO}$ terminated systems. However, it is still important to study the nature of the 
electronic states in such systems since it is well known that the electronic properties crucially depend on 
the termination layer (\cite{sto-2003},\cite{sro-sto}). In this line, we have studied the electronic properties of 
$\rm{BiO}$ terminated TBFO surfaces in this section. 
\begin{figure*}[htbp!]
\centering
	\includegraphics[width=0.28\textwidth,height=0.26\textwidth]{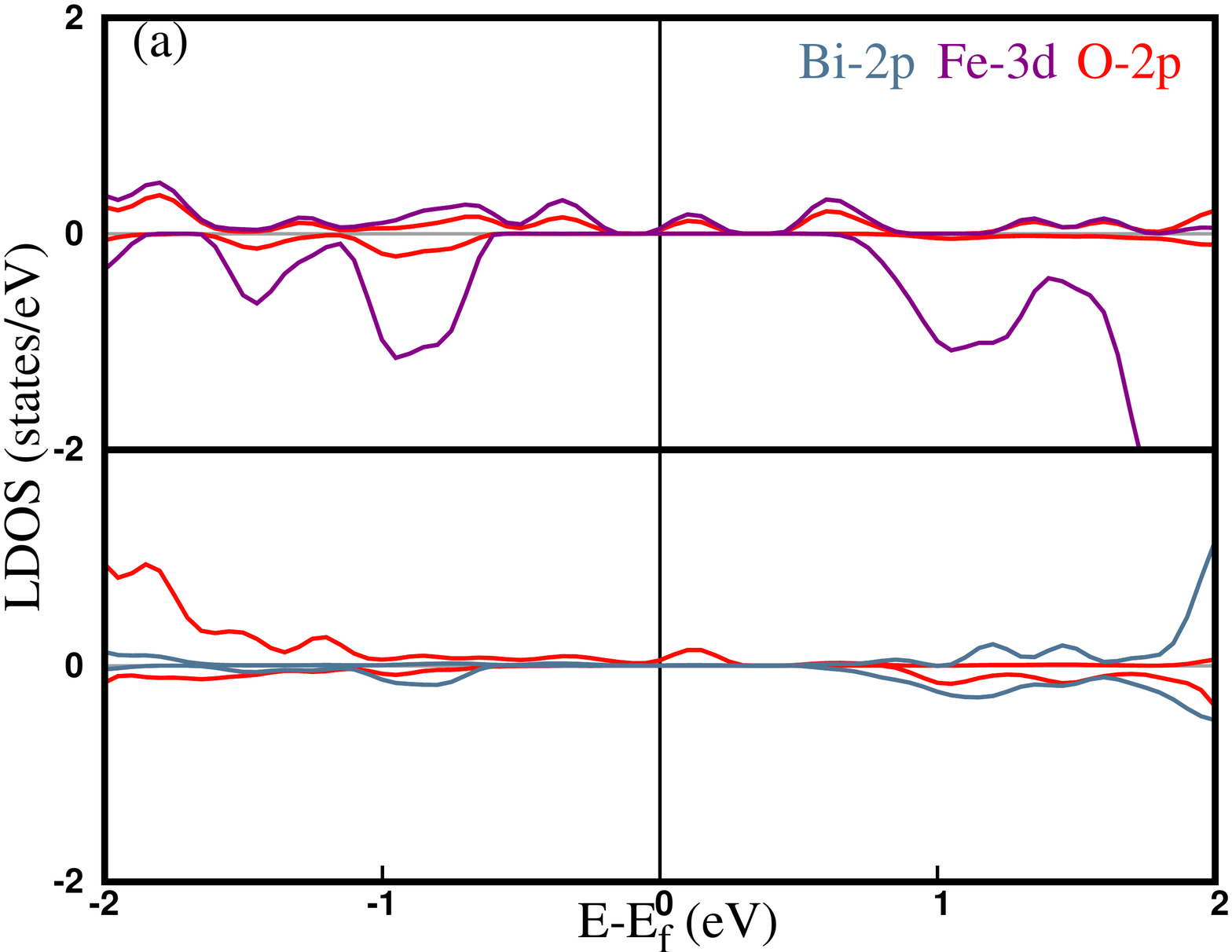}\hspace{0.6cm}
	\includegraphics[width=0.28\textwidth,height=0.26\textwidth]{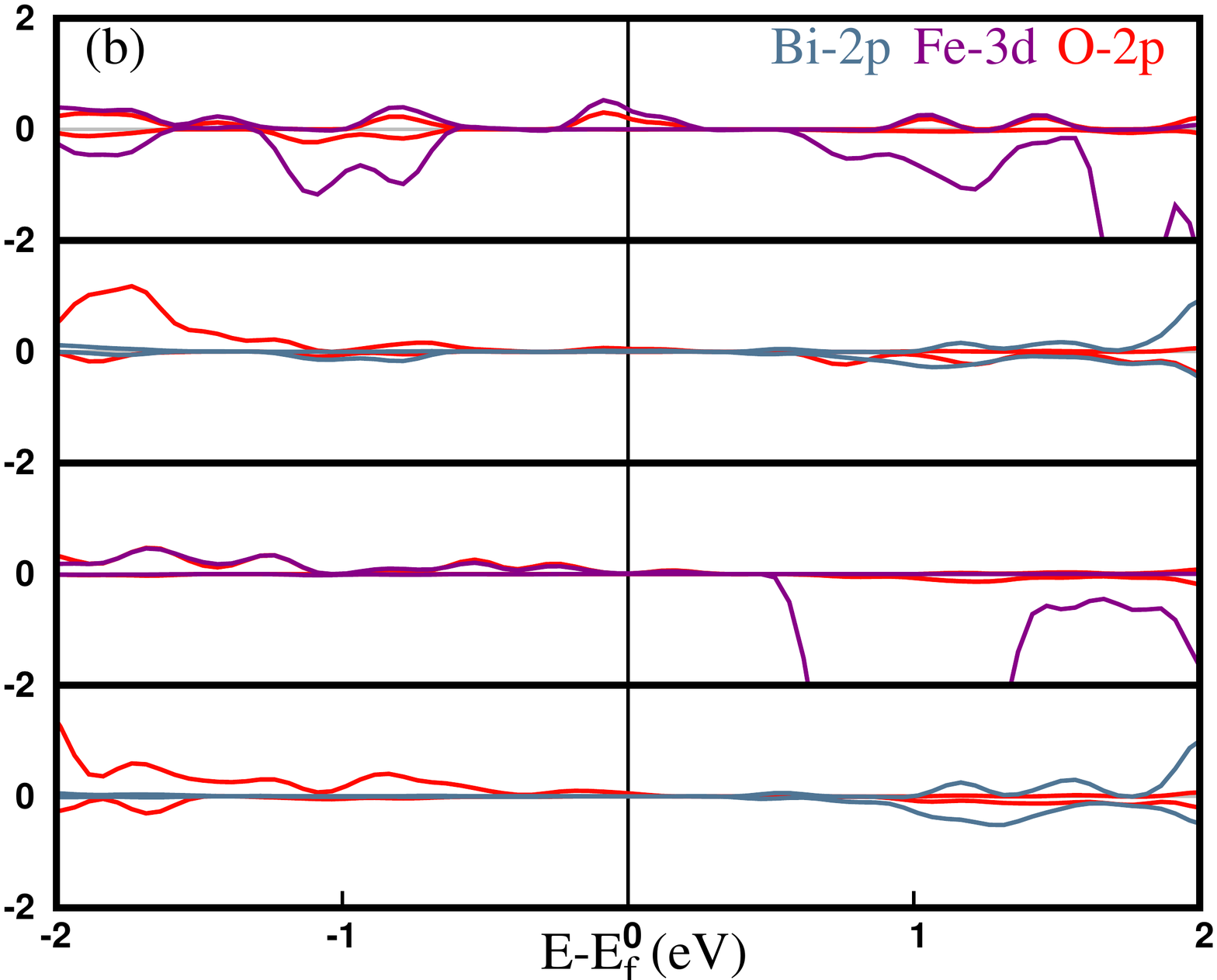}\hspace{0.6cm}
	\includegraphics[width=0.28\textwidth,height=0.26\textwidth]{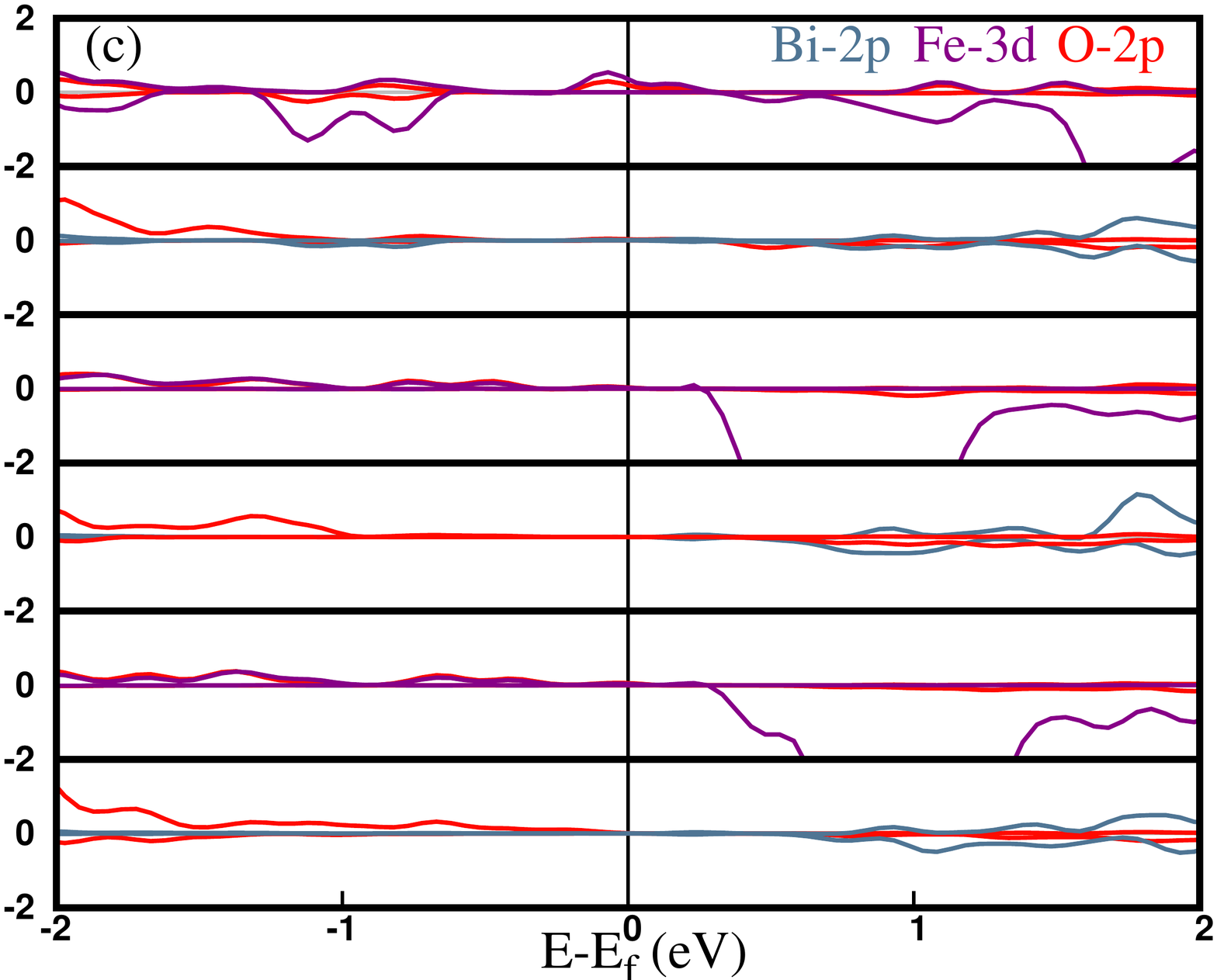}\hspace{0.6cm}
\vspace{-0.4cm}
\caption{Local density of states (LDOS) of structure-II with $\rm{FeO_2}$ termination (a) 1ml, (b) 2ml and (c) 3ml respectively.}
\label{Fig:st2-ldos-feo2}
\end{figure*}
Similar results are found for all the slabs with different thickness in all the structures. Here, only the results for 1ml thickness is presented.
In Fig~\ref{Fig:dos-bio-1l}, the total density of states(DOS) results for all the structures are presented. Contributions to the DOS from 
both the $\rm{BiO}$ and $\rm{FeO_2}$ layers are represented  separately. Substantial amount of contribution is observed 
towards the DOS from both the layers. From the DOS, it is evident that at the Fermi energy there are contributions 
from both the spin channels for all the structures resulting in metallic states at the surface having finite amount of total 
magnetization. 
It is important to note that in an earlier work \cite{bulk-paper} the bulk system was found to host 
wide range of electronic states for these four different structures. In the bulk, the structure-III and structure-IV were found to 
be in ferromagnetic metallic phase, while structure-I was a magnetic semiconductor and structure-II showed half-metallic character. 
However, our current results point towards dramatic change in the electronic properties with reduced dimensionality. 
In the next section we are going to see that a much richer electronic phases akin to the bulk system can appear when the slabs 
are $\rm{FeO_2}$ terminated. 
\begin{table}[htbp!]
\caption{Electronic properties for the $\rm{FeO_2}$ surface termination for different thickness of surface layers.}
\begin{center}
\begin{tabular}{c c c c c c c }
\hline
Structures & & 1ml  && 2ml && 3ml    \\
\hline 
I       && 2DHG,HMFM && 2DHG,HMFM && 2DHG,HMFM \\
II      && 2DHG,HMFM && 2DHG,HMFM && 2DHG,HMFM\\
III     && 2DHG,HMFM && 2DHG && Metal  \\
IV      && 2DHG,HMFM && 2DHG && Metal  \\
\hline
\label{Table:2deg-hm}
\end{tabular}
\end{center}
\end{table}
\subsubsection{$\rm{FeO_2}~termination$}\label{feo2-termination}
In this section, the results for asymmetrical slab geometries with $\rm{FeO_2}$ terminated top layer are presented. All the 
results are summed up in the Table.~\ref{Table:2deg-hm}.
We are going to demonstrate that in this case the electronic properties change dramatically as compared with the  $\rm{BiO}$ terminated 
slab. Furthermore, the electronic properties have been found to be very sensitive to the thickness of the surface layers, except for
structure-I. We have performed a systematic analysis of the surface electronic states. These results are presented in 
Figs.~\ref{Fig:1l-dos-feo2}, \ref{Fig:2l-dos-feo2} and \ref{Fig:3l-dos-feo2}. 
In Fig.~\ref{Fig:1l-dos-feo2}, the contribution of each atomic layer towards the DOS for a 
slab with 1ml thick surface are presented. It can be observed that in this case, each structure shows a weak but clear signature of 
half-metallic states at the surface. Furthermore, both the  $\rm{FeO_2}$ and $\rm{BiO}$ layers contribute hole type charge 
carriers at the Fermi energy. However, there is a crucial 
difference between structure-II and all the other structures i.e., it is found to be the most stable structure as we have 
seen in the thermodynamic analysis~\ref{thermodynamic stability}.
This indicates that structure-II is the most suitable candidate to explore the possibility of the existence of the 
2DHG in this system experimentally. To ascertain if 2DHG persists to higher slab thickness, slab structures with increasing surface mono-layers 
of 2ml and 3ml also have been studied. The layered DOS for 2ml and 3ml thick surfaces have been 
presented in Figs.~\ref{Fig:2l-dos-feo2} and ~\ref{Fig:3l-dos-feo2} respectively. 
It can be observed from Figs.~\ref{Fig:2l-dos-feo2}(b) and ~\ref{Fig:3l-dos-feo2}(b) that the surface of
structure-II yet behaves as HMFM. Furthermore, the charge carriers are still hole type. 
On the other hand, for structure III and IV the 2DHG survives only upto 2ml thick surface. But these surfaces no more 
behave as HMFM.
However, on further increase of the surface layers to 3ml, the surfaces of 
these structures become metallic as can be seen from Figs.~\ref{Fig:3l-dos-feo2}(c) and (d). It is important to note that like 
structure II, structure-I also hosts half-metallic surface states having hole-type charge carriers 
all the way upto 3ml thick surface. However, as we have found from our thermodynamic analysis this 
structure has a tendency to be unstable as the slab thickness grows beyond 1ml, it may not be very 
suitable to study it experimentally.

\begin{figure*}[htbp!]
\centering
	\includegraphics[width=0.32\textwidth,height=0.25\textwidth]{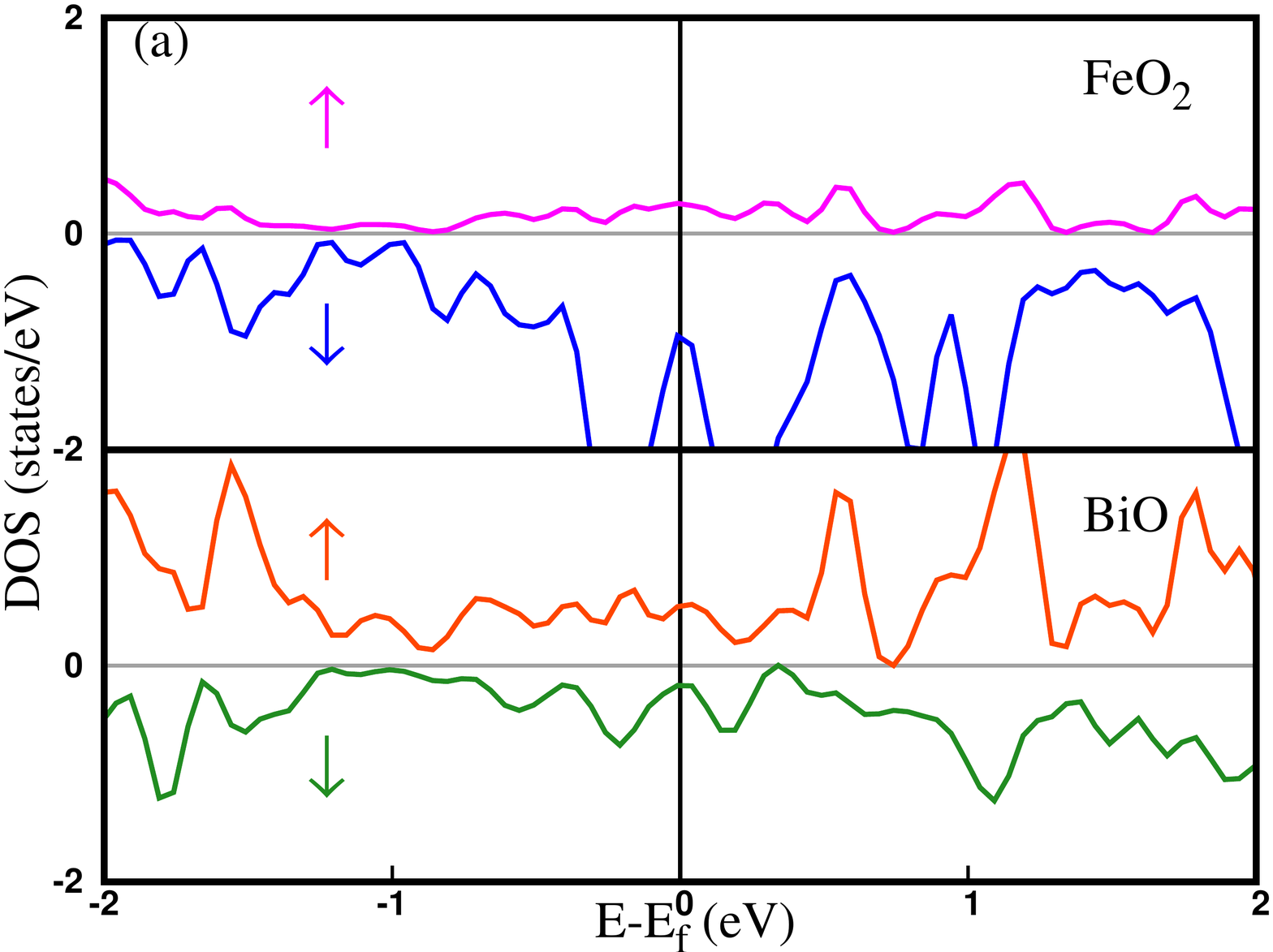}\hspace{-0.1cm}
	\includegraphics[width=0.32\textwidth,height=0.25\textwidth]{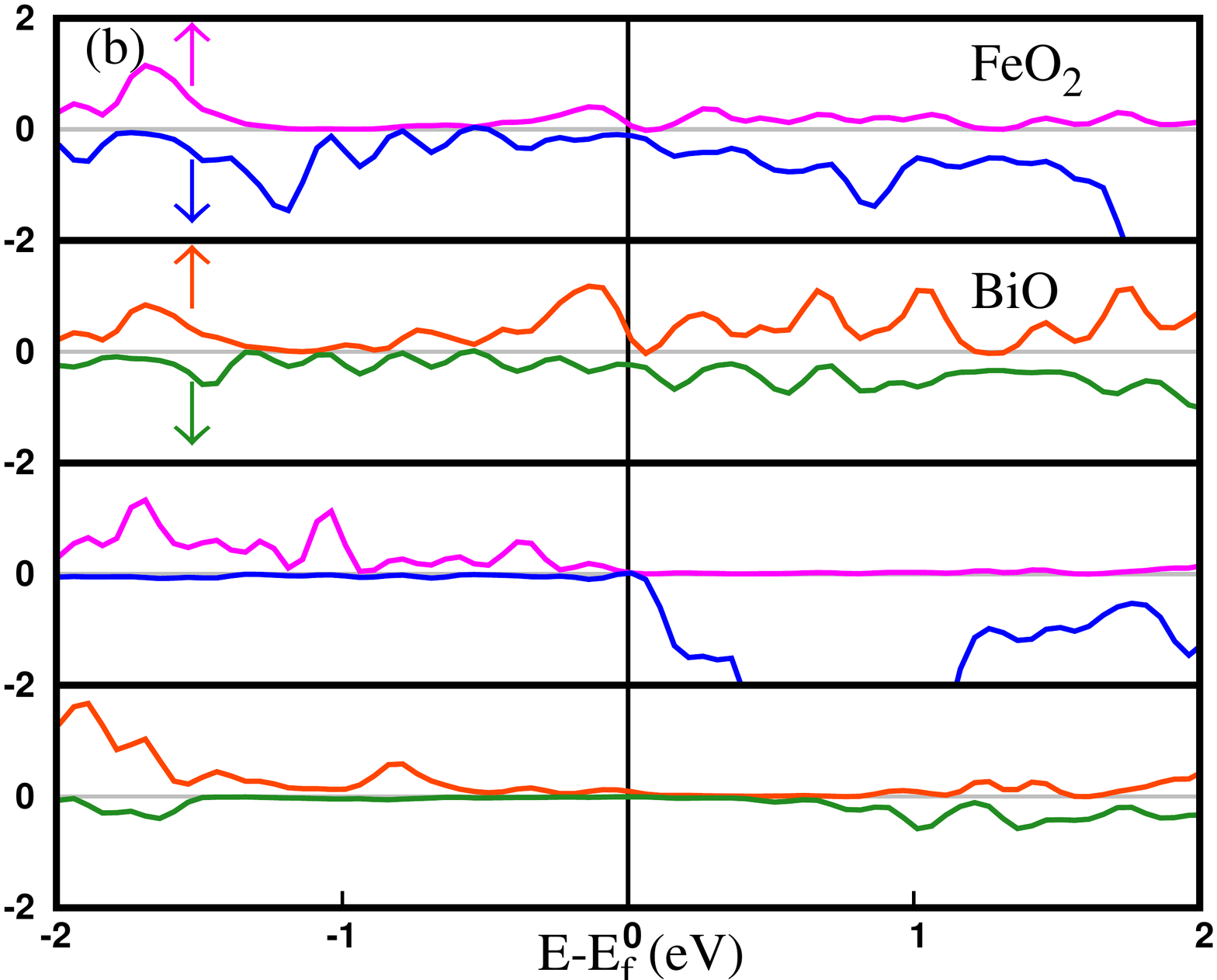}\hspace{-0.1cm}
	\includegraphics[width=0.32\textwidth,height=0.25\textwidth]{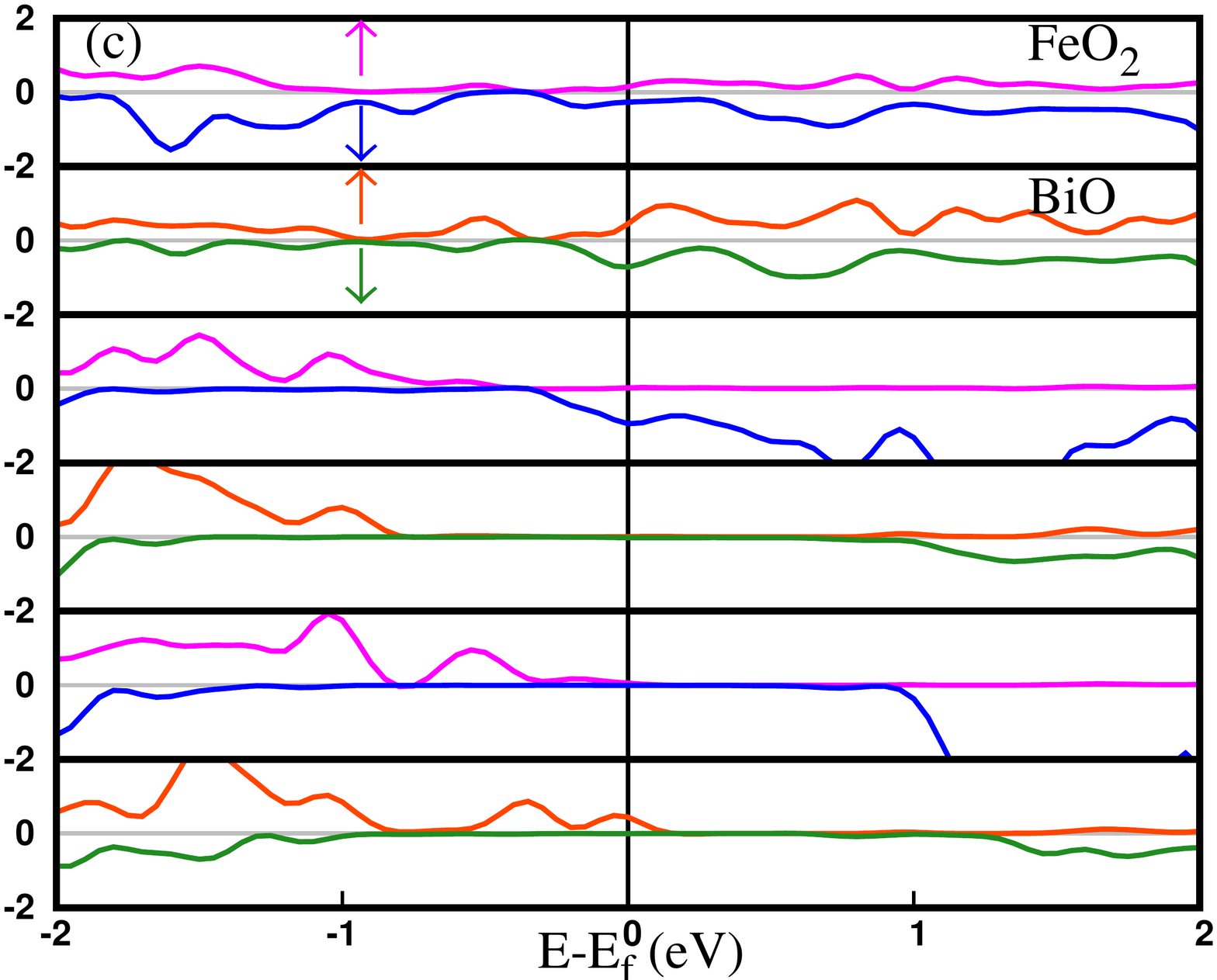}
\vspace{-0.4cm}
\caption{Layered density of states (DOS) corresponding to Structure-II with oxygen vacancy at `V1' site  (a) 1ml, (b) 2ml and (c) 3ml respectively.}
\label{Fig:st2-ov-dos-s1}
\end{figure*}

\begin{figure*}[htbp!]
\centering
	\includegraphics[width=0.32\textwidth,height=0.25\textwidth]{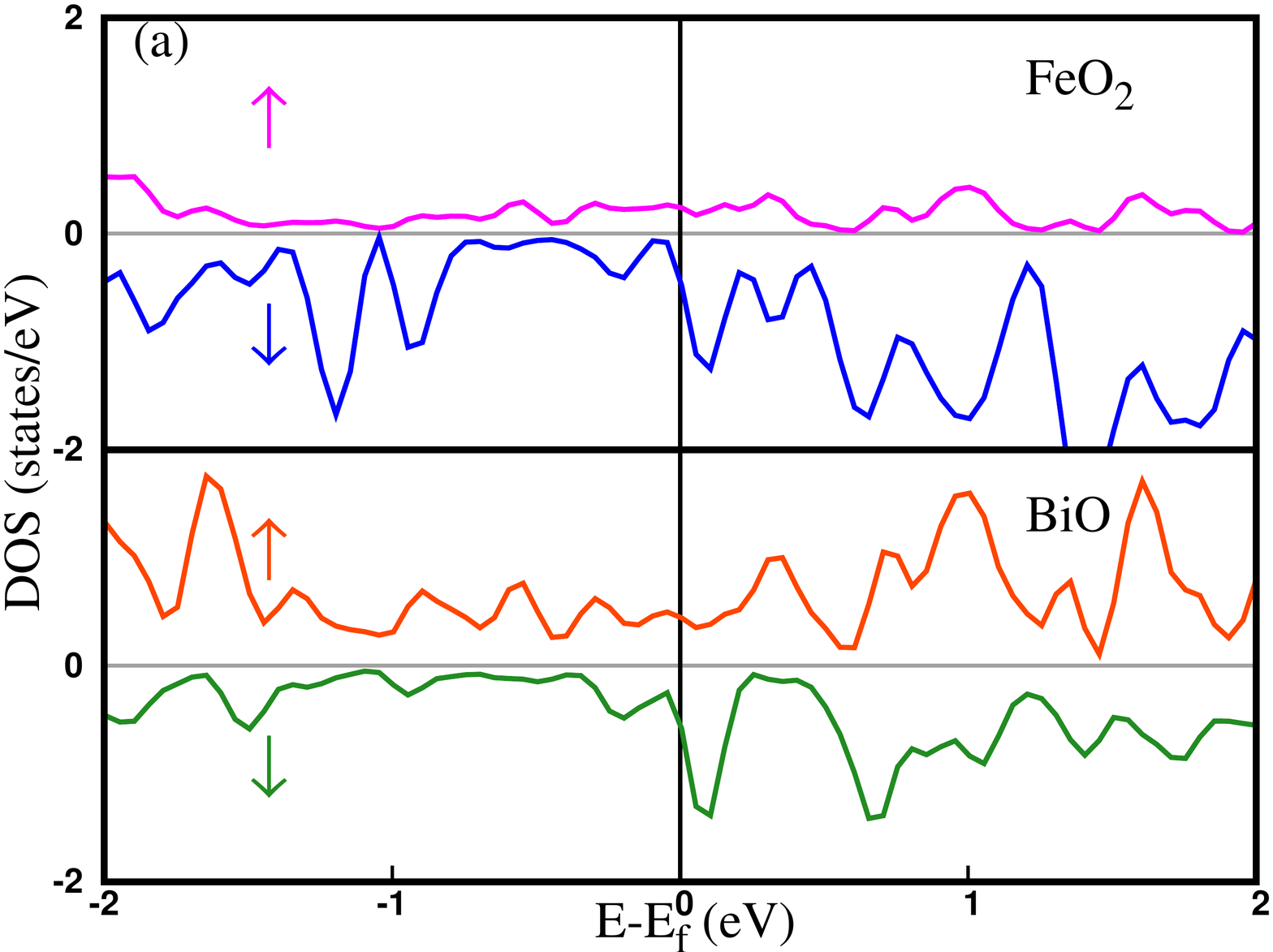}\hspace{-0.1cm}
	\includegraphics[width=0.32\textwidth,height=0.25\textwidth]{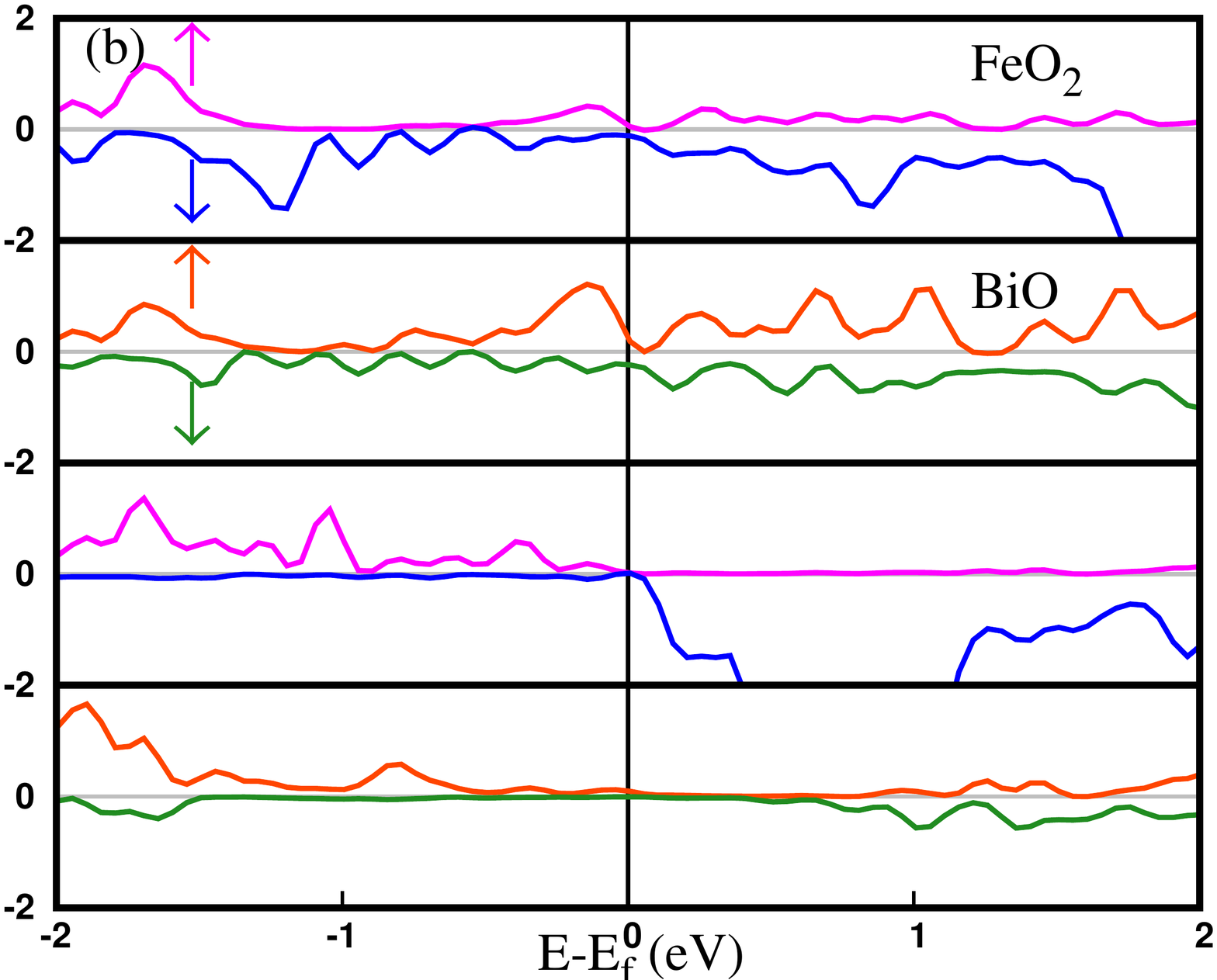}\hspace{-0.1cm}
	\includegraphics[width=0.32\textwidth,height=0.25\textwidth]{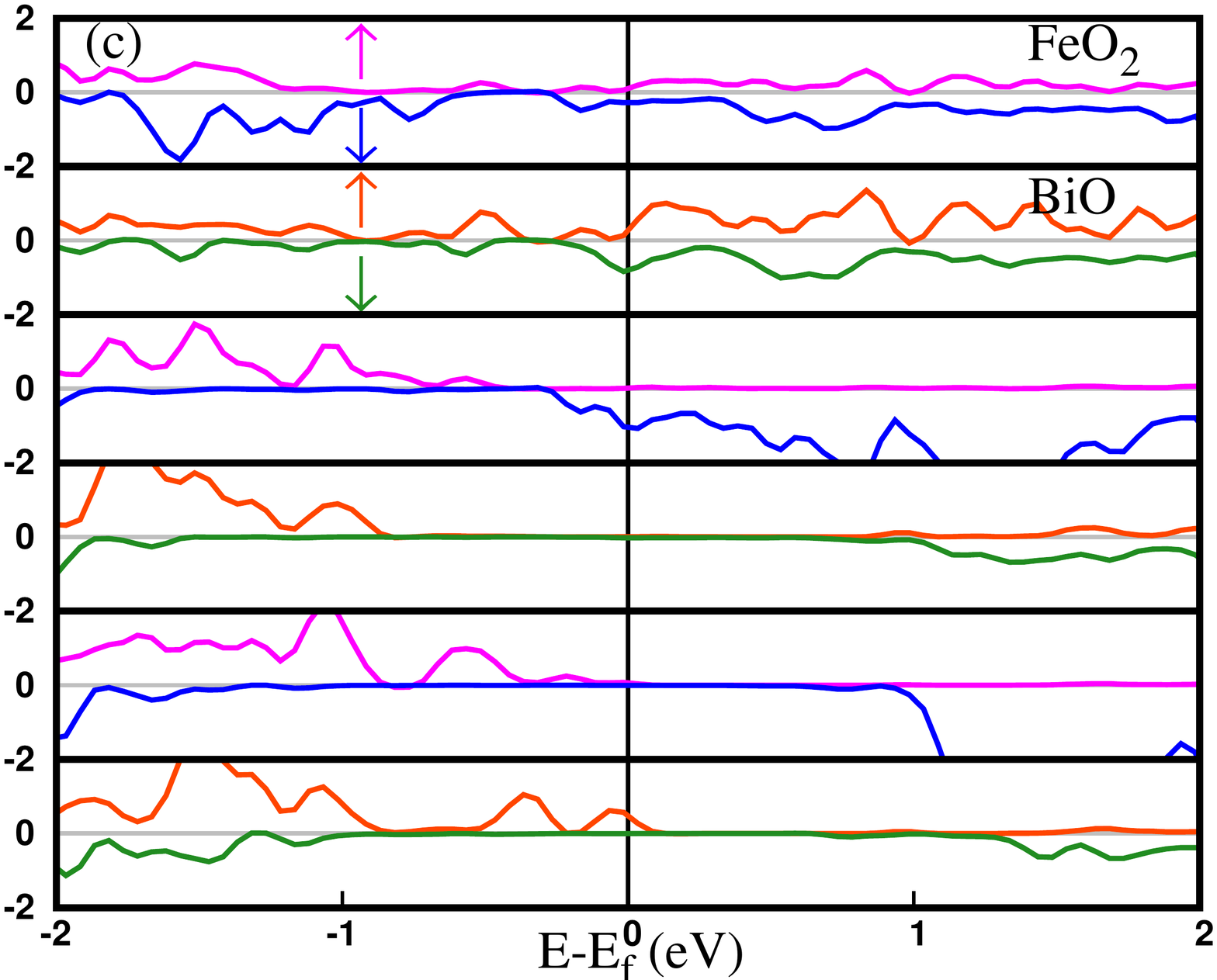}
\vspace{-0.4cm}
\caption{Layered density of states (DOS) corresponding to Structure-II with oxygen vacancy at `V2' site  (a) 1ml, (b) 2ml and (c) 3ml respectively.}
\label{Fig:st2-ov-dos-s2}
\end{figure*}
\begin{table}[htbp!]
\caption{\label{Table-1:} 
{For the structure-I, all the parameters are given in the Table. 
Top row: The angle $\rm{O_e-Fe-O_e}$ (in deg), where $\textrm{O}_{e}$ is the 
equatorial oxygen O2/O3. Second row: The angle $\rm{O1-Fe-O1}$ (in deg), where O1 is the 
axial oxygen. Third row: the bond-length O1-Fe (in unit \AA{}). Bottom row: Fe-$\textrm{O}_{e}$ bond-length 
(in \AA{}).}}
\begin{center}
\begin{tabular}{c c c c c c c c }
\hline
 &  & Fl    & 1ml  & 2ml & 3ml   \\
\hline
$\rm{O_e-Fe-O_e}$($\theta_1$)    &  & 146.85    & 157.50 & 158.59  & 158.70 \\
$\rm{O1-Fe-O1}$($\theta_2$)    &  & 180.00    & 180.00  & 180.00  & 180.00 \\
O1-Fe       &  & 2.23   & 2.73   & 2.67   &  2.68 \\
$\textrm{O}_{e}$-Fe      &  & 1.91     & 1.87  & 1.87   &  1.87  \\
\hline
{\label{Table:str-I}}
\end{tabular}
\end{center}
\end{table}
As structure-II is the most stable and preferable structure to search for 
2DHG henceforth, this structure is inspected in more detail. To further analyze the orbital 
contributions towards the 2DHG in this structure, we have computed the local density of states (LDOS). 
The LDOS results for this structure are presented in Fig.~\ref{Fig:st2-ldos-feo2}.
The results for alternative layers i.e., $\rm{FeO_2}$ and $\rm{BiO}$ have been presented separately
to understand the orbital contributions from individual layer. For 1ml thick surface the contributions of 
Fe-3d and O-2p orbitals from the $\rm{FeO_2}$ plane and O-2p contribution from the $\rm{BiO}$ layer can be observed 
from Fig.~\ref{Fig:st2-ldos-feo2}(a). Similarly, in case of 2ml and 3ml thick surfaces, the contributions 
towards the LDOS comes solely from the Fe-3d and O-2p orbitals of the surface layers. In addition all the contributions are found from 
the spin-up component. In general, we have found that Bi does not contribute any states at the surface of any structure. 
Furthermore the equivalent contributions are found in all other cases except for the 3ml thick surfaces belong to structure-III and 
IV. The LDOS results for these surfaces have been discussed in Appendix~\ref{Appendix:ldos}.

 We have also investigated the $\rm{FeO_2}$ terminated stoichiometric slabs with
full ionic relaxation i.e., without keeping any fixed atomic layers at its bulk atomic position, in order 
to inquire the presence of the 2DHG at the surface in the full slabs. For this study, the 2ml and 4ml slabs are taken 
into account. These results are presented in  Appendix~\ref{Appendix:full-slab}.

 \begin{table}[htbp!]
\caption{\label{Table-2:} 
{Measured parameters for structure-II} similar to Table \ref{Table:str-I}.}
\begin{center}
\begin{tabular}{c c c c c c c c }
\hline
 &  & Fl    & 1ml  & 2ml & 3ml   \\
\hline
$\rm{O_e-Fe-O_e}$($\theta_1$)    &  & 147.70   & 166.34 & 168.17  & 171.43 \\
$\rm{O1-Fe-O1}$($\theta_2$)    &  & 180.00    & 180.00  & 180.00  & 180.00 \\
O1-Fe       &  & 2.30   & 2.46   & 2.37  &  2.27 \\
$\textrm{O}_{e}$-Fe      &  &1.96    & 1.90  & 1.90   &  1.89  \\
\hline
{\label{Table:str-II}}
\end{tabular}
\end{center}
\end{table}
Moreover, in order to understand the intricate connection between the structural and electronic properties
of the surfaces, we now present the variation in structural parameters e.g., O-Fe-O angle and the O-Fe
distance for all the structures.
The top surface layer for all the thicknesses are studied in Table~\ref{Table:str-I},~\ref{Table:str-II},~\ref{Table:str-III} and 
\ref{Table:str-IV} belong to structure-I, structure-II, structure-III and structure-IV respectively. 
In Fig.~\ref{Fig:cell-structure} the equatorial and the axial oxygens are labeled as $\rm{O_e}$ and $\rm{O1}$ 
respectively. The angle $\rm{O_e-Fe-O_e}$ and $\rm{O1-Fe-O1}$ are named as $\theta_1$ and $\theta_2$ respectively. 
The $\theta_2$ is found to be same for all the cases i.e. 180\degree, whereas there is a substantial difference in 
$\theta_1$ for the top surface as compared to the fixed layers (Fl). From the Tables, it can be seen that in general 
$\theta_1$ has a higher value at the top surface compared to the fixed layers. In addition $\theta_1$ 
also varies from structure to structure. The highest difference has been found in the case of structure-II. 
The $\theta_1$ for the top layer remains almost constant with the increase of surface thickness 
in the case of structure-I, III and IV, while in the case of structure-II it increases monotonically. 
\begin{table}[htbp!]
\caption{\label{Table-3:} 
{Measured parameters for structure-III} similar to Table \ref{Table:str-I}.}
\begin{center}
\begin{tabular}{c c c c c c c c }
\hline
 &  & Fl   & 1ml  & 2ml & 3ml   \\
\hline
$\rm{O_e-Fe-O_e}$($\theta_1$)    &  &161.19   & 178.78 & 178.28  & 178.28 \\
$\rm{O1-Fe-O1}$($\theta_2$)    &  & 180.00    & 180.00  & 180.00  & 180.00 \\
O1-Fe       &  & 2.11  & 1.94  & 1.98   &  1.97 \\
$\textrm{O}_{e}$-Fe & & 1.97  & 1.94    & 1.94  & 1.94    \\
\hline
{\label{Table:str-III}}
\end{tabular}
\end{center}
\end{table}
The $\rm{O_e-Fe}$ distance at the top layer shows a behavior that is common to all the structures that is,
it is lower compared to the fixed layer and it does not change with the increase of the thickness of the surface 
layers. However, the $\rm{Fe-O1}$ distance at the top surface for structure-I and structure-II has a completely 
opposite tendency in comparison to the structures III and IV. From Tables \ref{Table:str-I} and \ref{Table:str-II}
we observe that at the top surface of the 1ml thick surface of the structures-I and II, the $\rm{O1-Fe}$ distance 
increases to a higher value compared to that in the fixed layers and then it decreases monotonically with further 
increase in the thickness of the surface layer. However, this distance still remains higher than that in the fixed 
layers. Interestingly, in the case of structures-III and IV the $\rm{O1-Fe}$ distance at the top of 1ml thick surface 
have been found to be lower than that in the fixed layer. It then increases slightly with the increase in the thickness
of the surface layer. However, the $\rm{O1-Fe}$ distance at the top surface always remains lower than the fixed layers.
Effectively top layer moves inward in z-direction. 
\begin{table}[htbp!]
\caption{\label{Table-4:} 
{Measured parameters for structure-IV} similar to Table \ref{Table:str-I}.}
\begin{center}
\begin{tabular}{c c c c c c c c }
\hline
 &  & Fl    & 1ml  & 2ml & 3ml   \\
\hline
$\rm{O_e-Fe-O_e}$($\theta_1$)    &  &162.67   & 179.91 & 179.32  & 179.56 \\
$\rm{O1-Fe-O1}$($\theta_2$)    &  & 180.00    & 180.00  & 180.00  &180.00 \\
O1-Fe     &  & 2.12 & 1.89   & 1.91   &  1.91 \\
$\textrm{O}_{e}$-Fe& & 1.99  & 1.97   & 1.97  & 1.97    \\
\hline
{\label{Table:str-IV}}
\end{tabular}
\end{center}
\end{table}

Next we proceed to discuss the results of magnetic moments of all the structures. As expected, the major contribution to magnetization is 
due to `Fe', which is similar to the bulk TBFO\cite{bulk-paper}. The `Fe' magnetic moments for each $\rm{FeO_2}$ plane of the 
surface layers have been presented in Table~\ref{Table:5}. In contrast to the variable nature of the structural parameters 
from one structure to the other, the magnetic moment shows an uniform behaviour in some aspect across all the structures. 
In all the cases, the magnetic moment of `Fe' at the outermost surface layer has been found to be the lowest. As we move 
 below the top layer, the magnetic moment increases and it remains almost same as we move towards the bottom layer.
However, there are still some interesting differences that can be observed from data presented in Table~\ref{Table:5}. 
For structure-I and II the magnetic moment of `Fe' at the top surface remains almost the same irrespective of the thickness 
of the surface layers, while in case of structure-III and IV the magnetic moment of `Fe' remains unchanged upto 2ml thick 
surface and then it suddenly increases for the 3ml thick surface. It is interesting to note that the 2DHG at the surface is 
destroyed in this case.
\begin{table}[htbp!]
\caption{\label{Table-5:} 
{The magnetic moment (M in $\mu_{B}$}) of `Fe' in the $\rm{FeO_2}$ plane.}
\begin{center}
\begin{tabular}{c c c c c c c ccccccc }
\hline
Structures   &&&& 1ml &&&& 2ml &&&& 3ml\\
\hline
Structure-I         &&&&   3.04 &&&&  3.04  &&&&   3.04  \\
                    &&&&       &&&&  3.65   &&&&   3.65  \\
                    &&&&       &&&&        &&&&  3.69  \\
\hline
Structure-II        &&&& 2.99 &&&& 2.97  &&&&  2.96  \\
                   &&&&    &&&& 3.72 &&&&  3.73            \\
                   &&&&    &&&&      &&&& 3.72 \\
\hline
Structure-III     &&&& 2.93 &&&& 2.94 &&&& 3.38 \\
             &&&&      &&&& 3.77&&&& 3.75 \\
             &&&&    &&&&     &&&& 3.79 \\
\hline
Structure-IV         &&&& 2.94 &&&& 2.94 &&&& 3.53  \\
               &&&&     &&&&  3.75 &&&&  3.77      \\
              &&&&      &&&&       &&&&  3.77  \\
\hline
\vspace{-0.5cm}
{\label{Table:5}}
\end{tabular}
\end{center}
\end{table}
\subsection{Oxygen vacancy}\label{oxygen-vacancy}
Since we have found that the $\rm{FeO_2}$ terminated structure-II is the most suitable candidate to study the 
existence of the 2DHG at the surface, in this section we have carried out a brief study of the effect of oxygen 
vacancy on the electronic properties of the surface states in this structure. In Fig.~\ref{Fig:ov} we have 
indicated the two inequivalent sites of oxygen at the top layer of a 2ml thick surface. The vacancies are generated at 
either of these two inequivalent positions which are named as `V1' and `V2'. In Figs.~\ref{Fig:st2-ov-dos-s1} and 
\ref{Fig:st2-ov-dos-s2} we have presented the individual layer resolved DOS for upto 3ml thick surface layers.  
It is evident from these results that having oxygen vacancy at the top surface immediately leads to the 
destruction of the 2DHG at the surface layers. 

\section{Conclusions}\label{conclusions}
In conclusions, motivated by the continued search for 2DEG/2DHG in polar perovskites, in this work we have primarily studied its
possibility at the surface of the  tetragonal $\rm{BiFeO_3}$(001) slabs in the presence of ferromagnetic ordering. 
We have studied the surface properties of asymmetrical slabs with the top surface either being $\rm{FeO_2}$ or $\rm{BiO}$ terminated. 
For an extensive analysis, we have studied four different structures with the each structure having the surface thickness upto 3ml. 
We have carefully carried out the thermodynamic analysis of each asymmetric slab. In general, we have found that the  $\rm{FeO_2}$ terminated 
asymmetrical slabs are thermodynamically more stable as compared to the $\rm{BiO}$ terminated slabs. In case of the $\rm{FeO_2}$ 
terminated asymmetrical slabs, structure-I turns out to be the most stable structure for a 1ml thick surface layer. However, 
structure-I becomes unstable with the increase in the thickness of the surface layers. Overall, we have found that the structure-II
is the most preferable structure when the thickness of the surface layer becomes larger than 1ml.

Following the thermodynamic analysis, we have investigated the nature of the electronic states at the surface. The $\rm{BiO}$ terminated 
surfaces have been found to be a simple ferromagnetic metal for all the structures, and for all the thicknesses. However, the 
$\rm{FeO_2}$ terminated surfaces show much more interesting and exciting electronic properties. For the surface of a 
1ml thick surface layer all the structures have been found to be half-metallic. Moreover, the charge carriers have been found to 
be of hole-type, effectively giving rise to a spin-polarized 2DHG at the surface. In case of structure-II, this spin-polarized 2DHG 
survives upto 3ml thick layers of the surface, while for structure-III and structure-IV it exists only upto 2ml thick surface.
Although, we have found that the 2DHG survives upto 3ml thick surface in the case of structure-I, we do not consider these 
results for further investigation as we have found this structure to be unstable beyond 1ml thick surface.
\begin{figure*}[htbp!]
\centering
	\includegraphics[width=0.32\textwidth,height=0.25\textwidth]{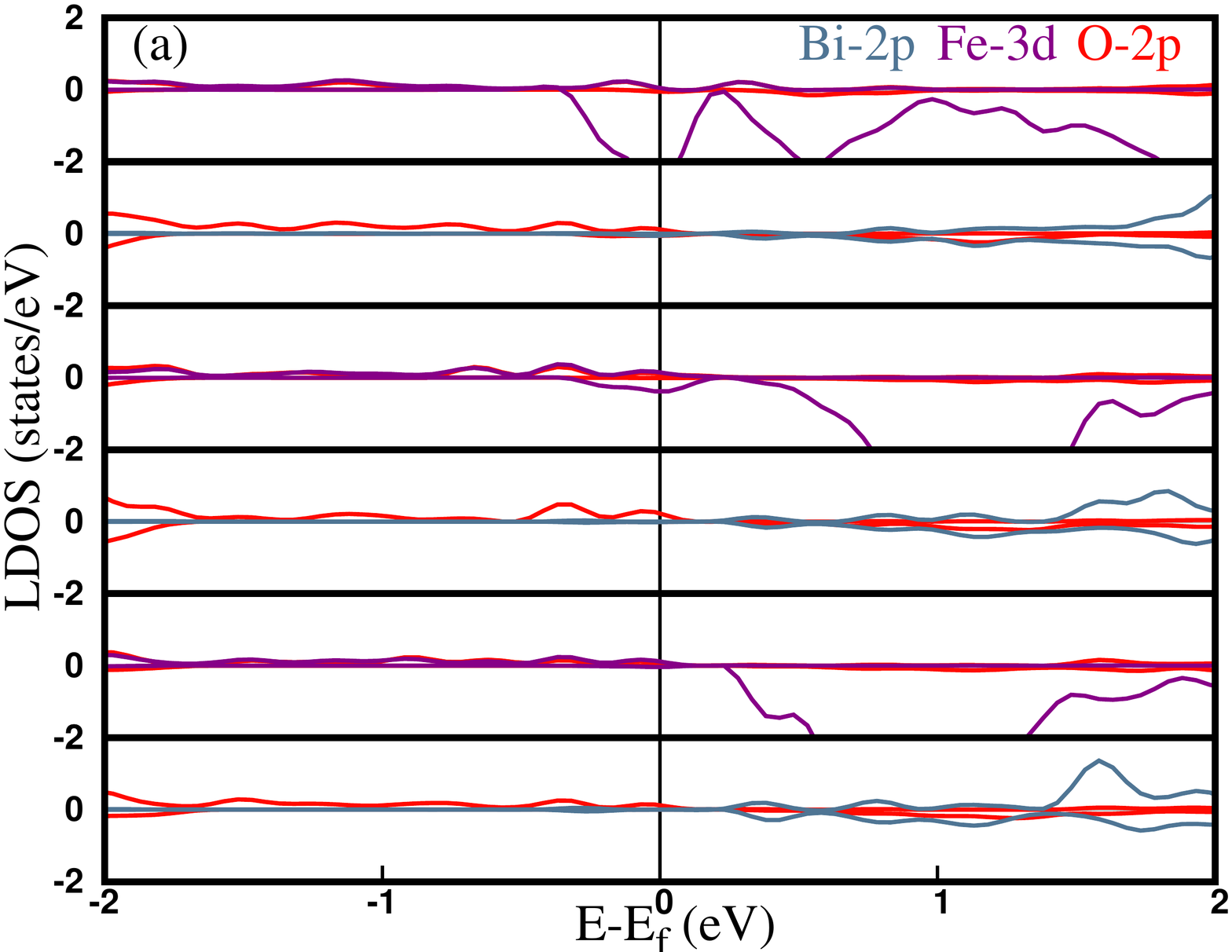}\hspace{0.5cm}
	\includegraphics[width=0.32\textwidth,height=0.25\textwidth]{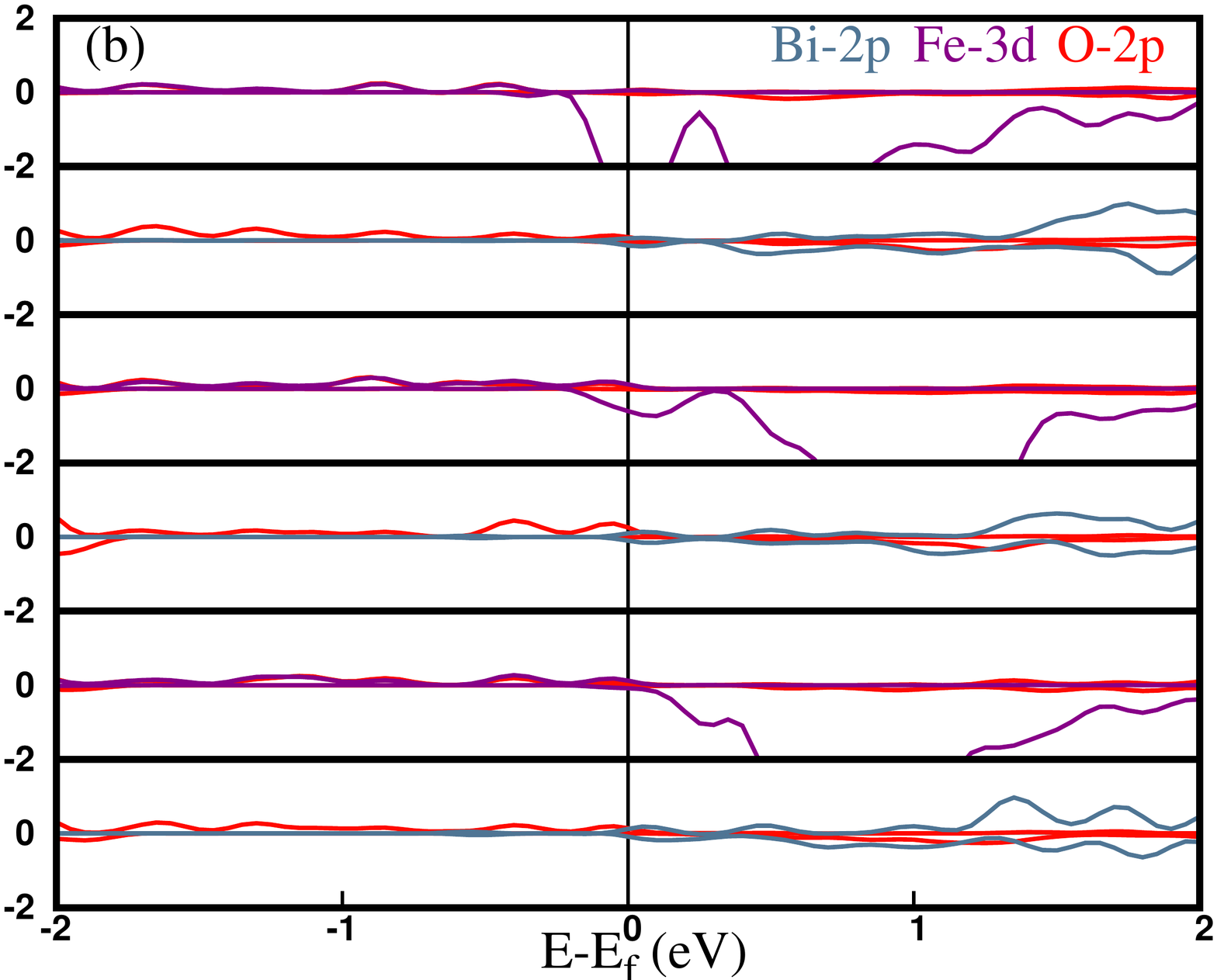}
\vspace{-0.4cm}
\caption{Local density of states (LDOS) of 3ml slab-thicknesses with $\rm{FeO_2}$ termination for  (a) structure-III and 
(b) structure-IV respectively.}
\label{Fig:3l-ldos-feo2}
\end{figure*}
From the thermodynamic analysis and the surface electronic states studies, structure-II turns out to possess the highest robustness and is 
demonstrated as suitable 
candidate to host 2DHG, and this structure has been investigated further. From the LDOS calculations, we have found that only 
`Fe' and 'O' contributes to the 2DHG, while `Bi' has no contribution towards it. Subsequently, we have investigated the fate of the 
2DHG in this structure in the presence of oxygen vacancy at the surface. We have found that oxygen vacancy leads to the 
destruction of the 2DHG, and the surfaces turn out to be metallic.
\section*{ACKNOWLEDGEMENTS}
 The authors 
would like to thank the computing facility provided by the National Institute of Technology, Rourkela and the 
computational facility provided by the Science and Engineering Research Board, Department of Science and Technology, 
India (Grant No: EMR/2015/001227).
S. J. is very much grateful to Sanchari Bhattacharya, NIT Rourkela, India for very useful discussions.
\appendix
\section{Local density of states (LDOS)}\label{Appendix:ldos}
The 2DHG is destroyed in the case of 3ml thick surfaces for structure-III and IV.
In order to understand the orbital contribution in these cases, we have studied the 
local density of states for  structure-III and structure-IV. The results are presented in the 
Fig.~\ref{Fig:3l-ldos-feo2}. In all the cases, along with 
the Fe-3d and O-2p orbital contributions the role of Bi-2p is also found.
\begin{figure*}[htbp!]
\centering
	\includegraphics[width=0.265\textwidth,height=0.25\textwidth]{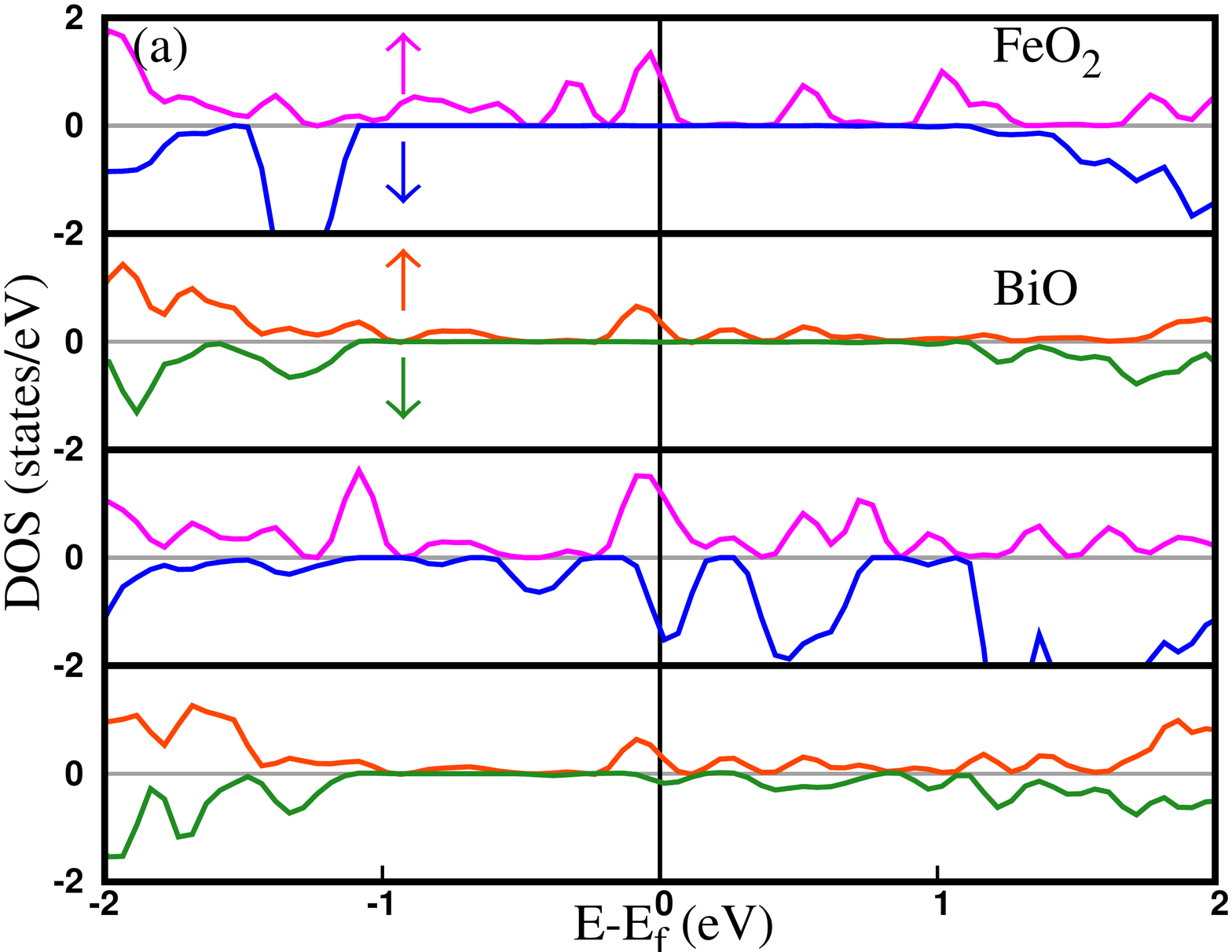}\hspace{-0.5cm}
	\includegraphics[width=0.265\textwidth,height=0.25\textwidth]{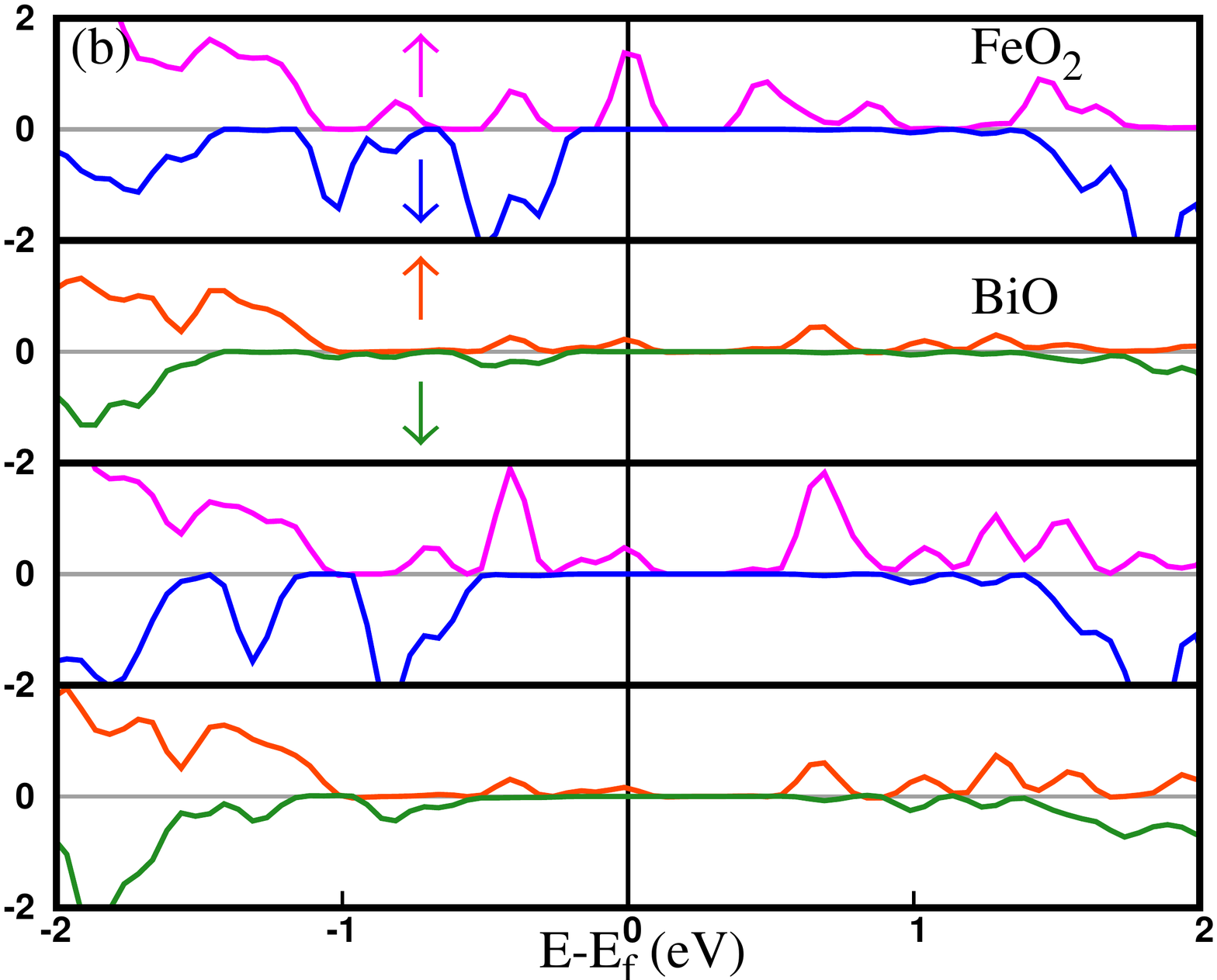}\hspace{-0.5cm}
	\includegraphics[width=0.265\textwidth,height=0.25\textwidth]{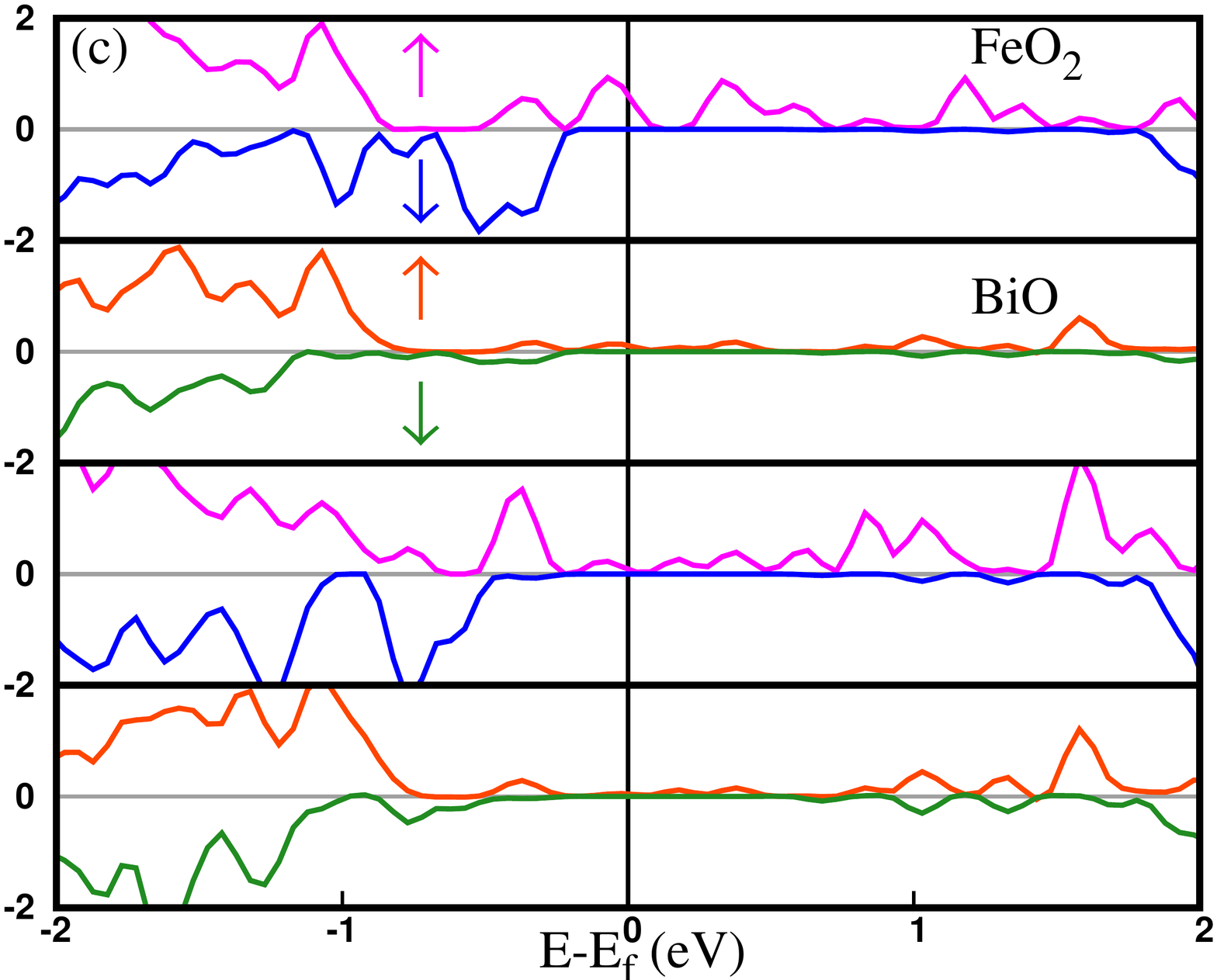}\hspace{-0.5cm}
	\includegraphics[width=0.265\textwidth,height=0.25\textwidth]{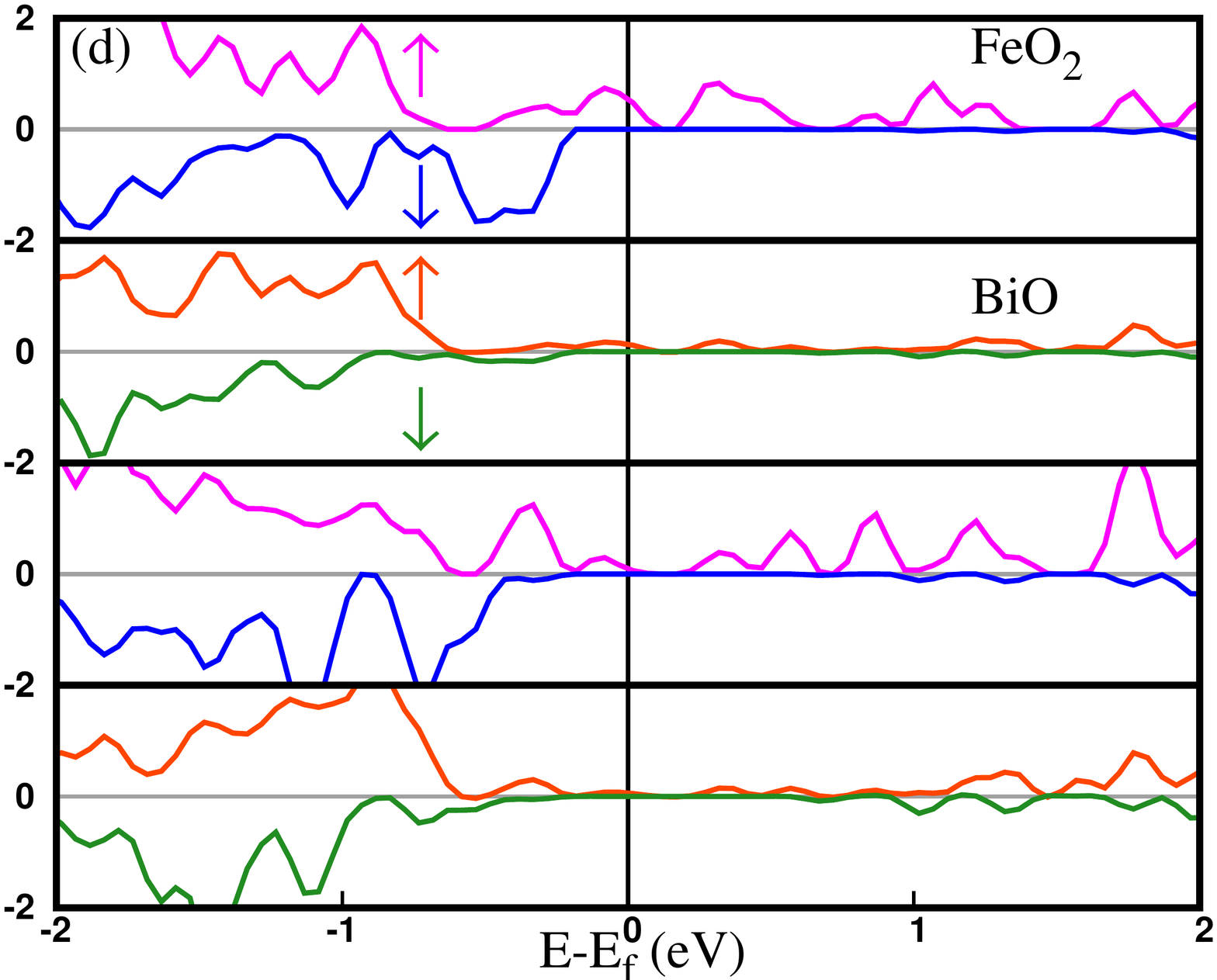}\hspace{-0.5cm}
\vspace{-0.4cm}
\caption{Density of states (DOS) of 2ml slab-thicknesses with $\rm{FeO_2}$ termination for (a) structure-I, (b) structure-II, 
(c) structure-III and (d) structure-IV respectively.} 
\label{Fig:fr-2l-dos-feo2}
\end{figure*}

\begin{figure*}[htbp!]
\centering
	\includegraphics[width=0.265\textwidth,height=0.25\textwidth]{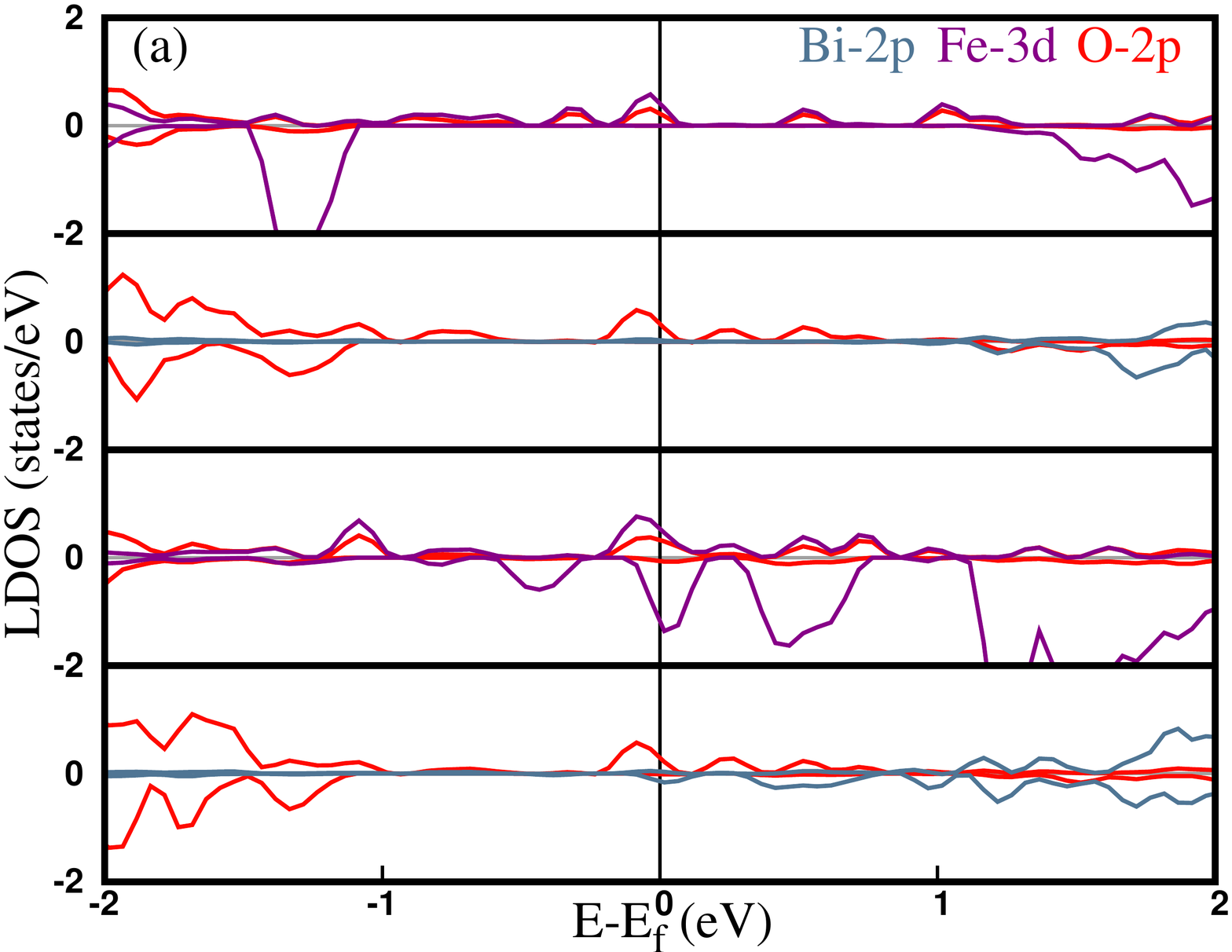}\hspace{-0.5cm}
	\includegraphics[width=0.265\textwidth,height=0.25\textwidth]{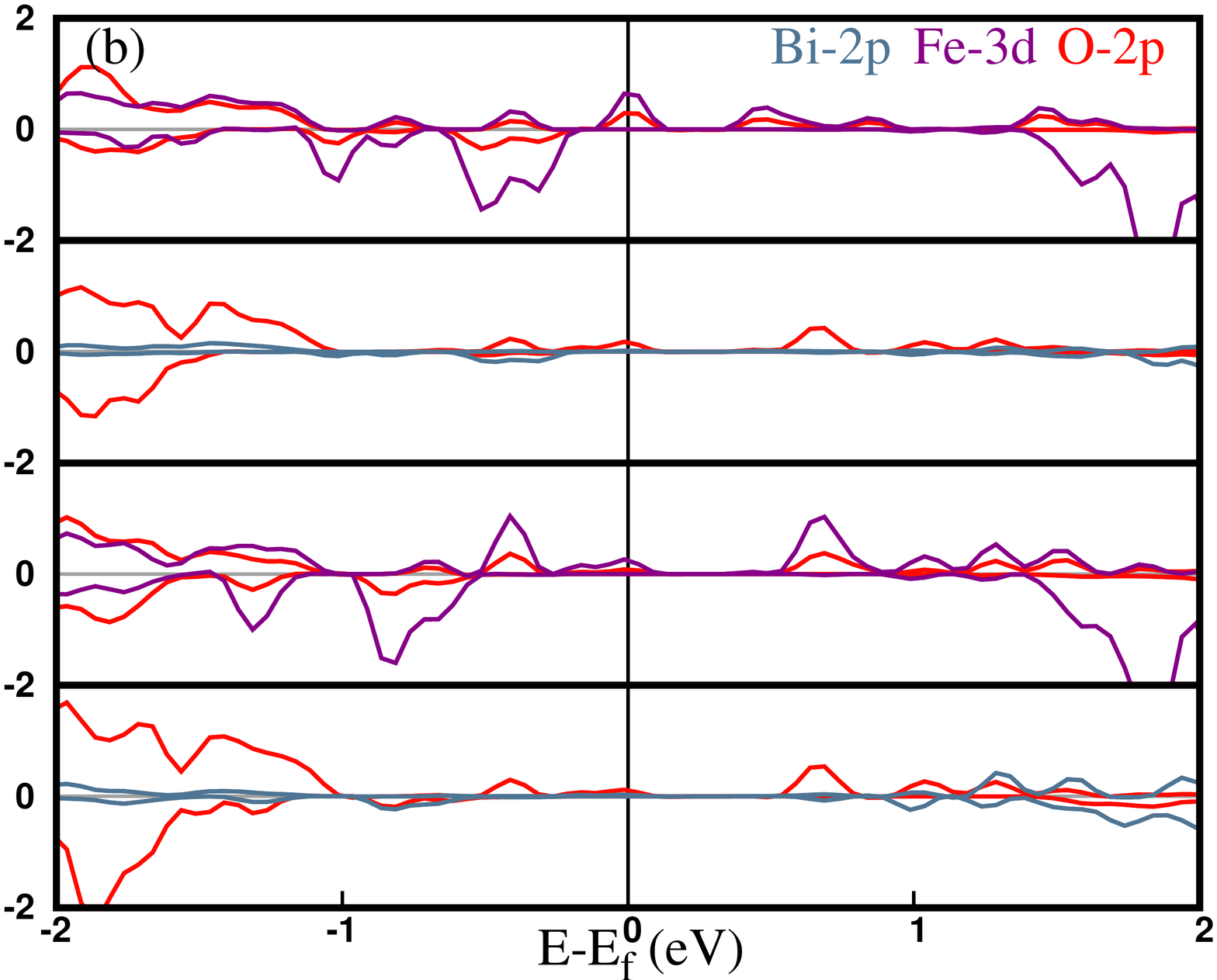}\hspace{-0.5cm}
	\includegraphics[width=0.265\textwidth,height=0.25\textwidth]{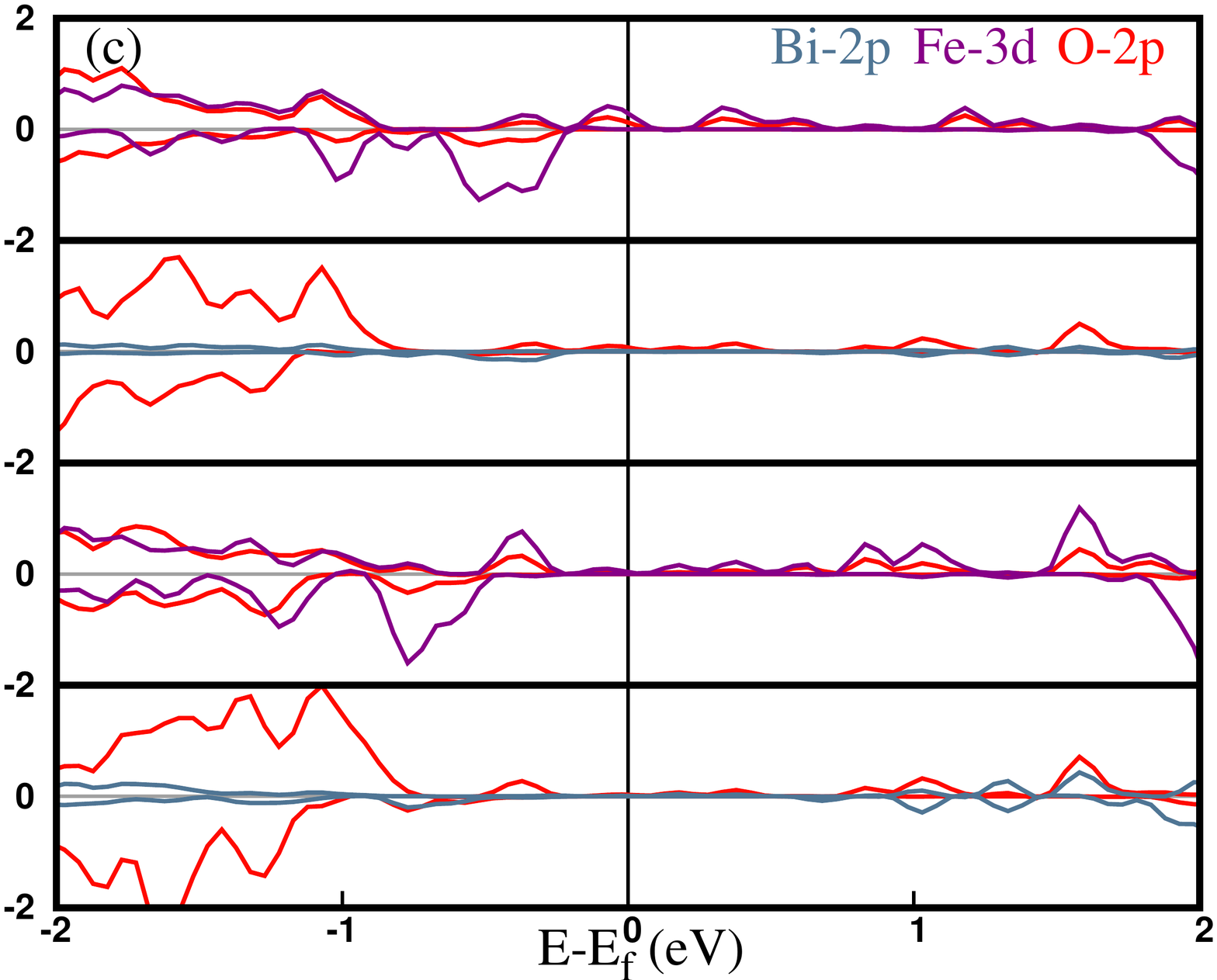}\hspace{-0.5cm}
	\includegraphics[width=0.265\textwidth,height=0.25\textwidth]{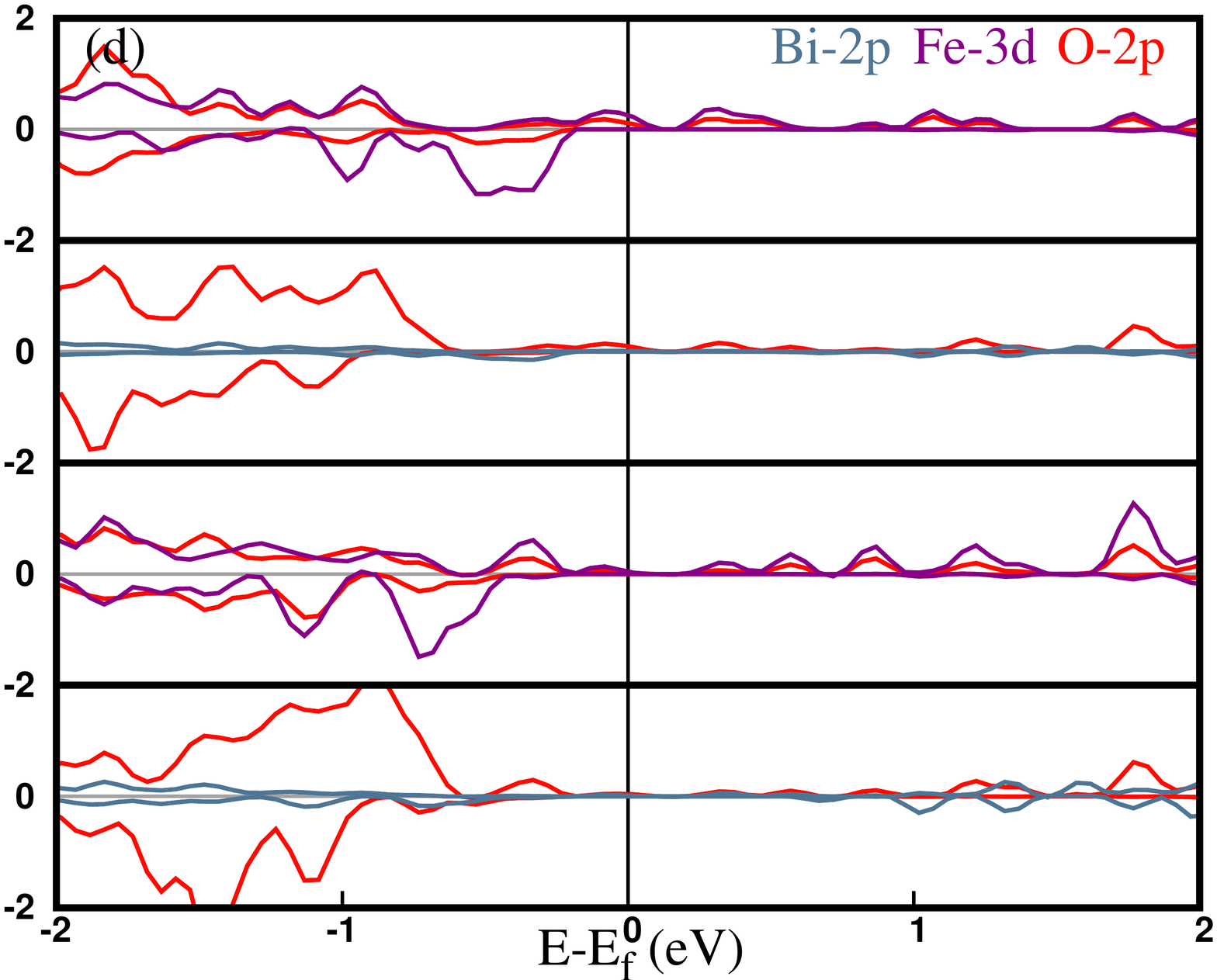}\hspace{-0.5cm}
\vspace{-0.4cm}
\caption{Local density of states (LDOS) of 2ml slab-thicknesses with $\rm{FeO_2}$ termination for (a) structure-I, (b) structure-II, 
(c) structure-III and (d) structure-IV respectively.} 
\label{Fig:fr-2l-ldos-feo2}
\end{figure*}
\begin{figure*}[htbp!]
\centering
	\includegraphics[width=0.265\textwidth,height=0.25\textwidth]{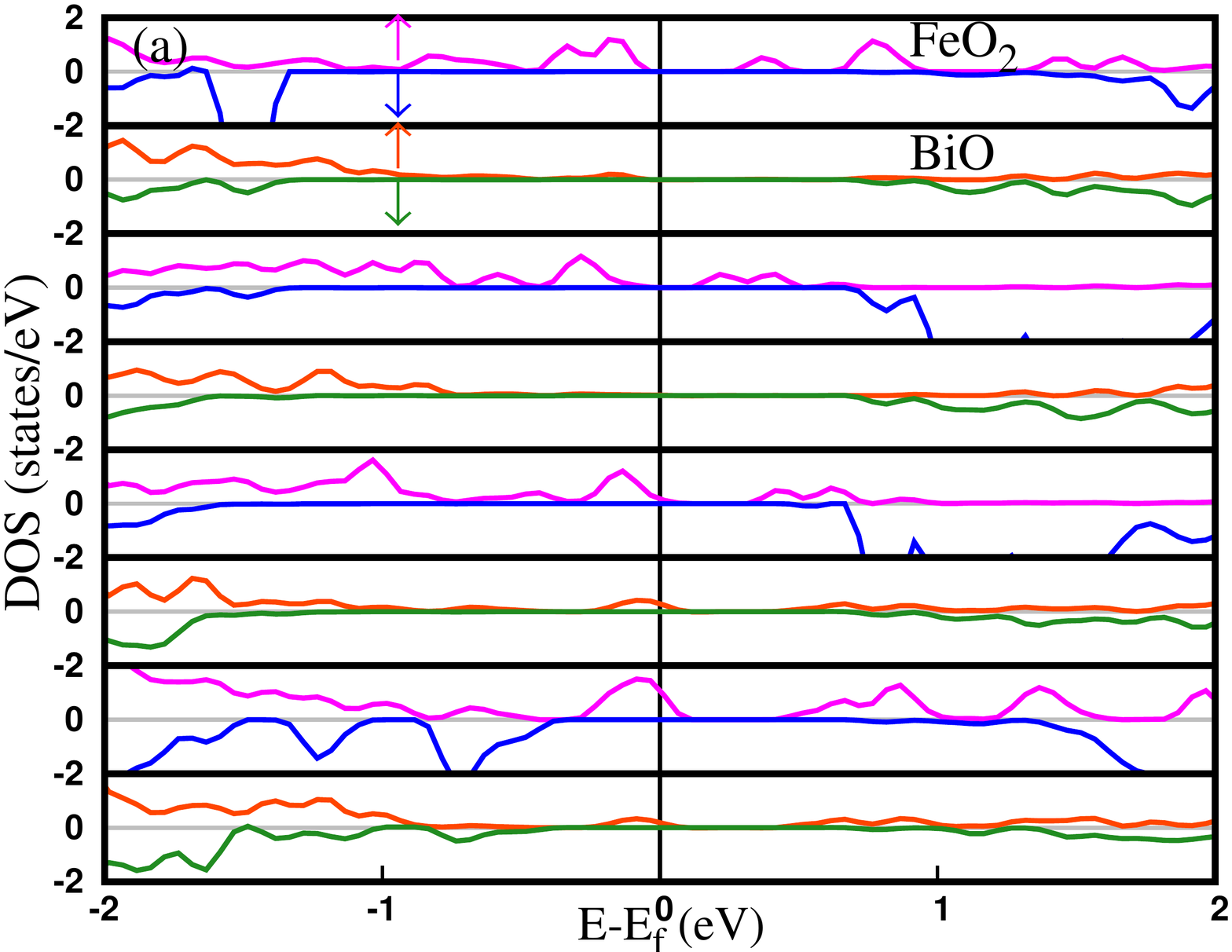}\hspace{-0.5cm}
	\includegraphics[width=0.265\textwidth,height=0.25\textwidth]{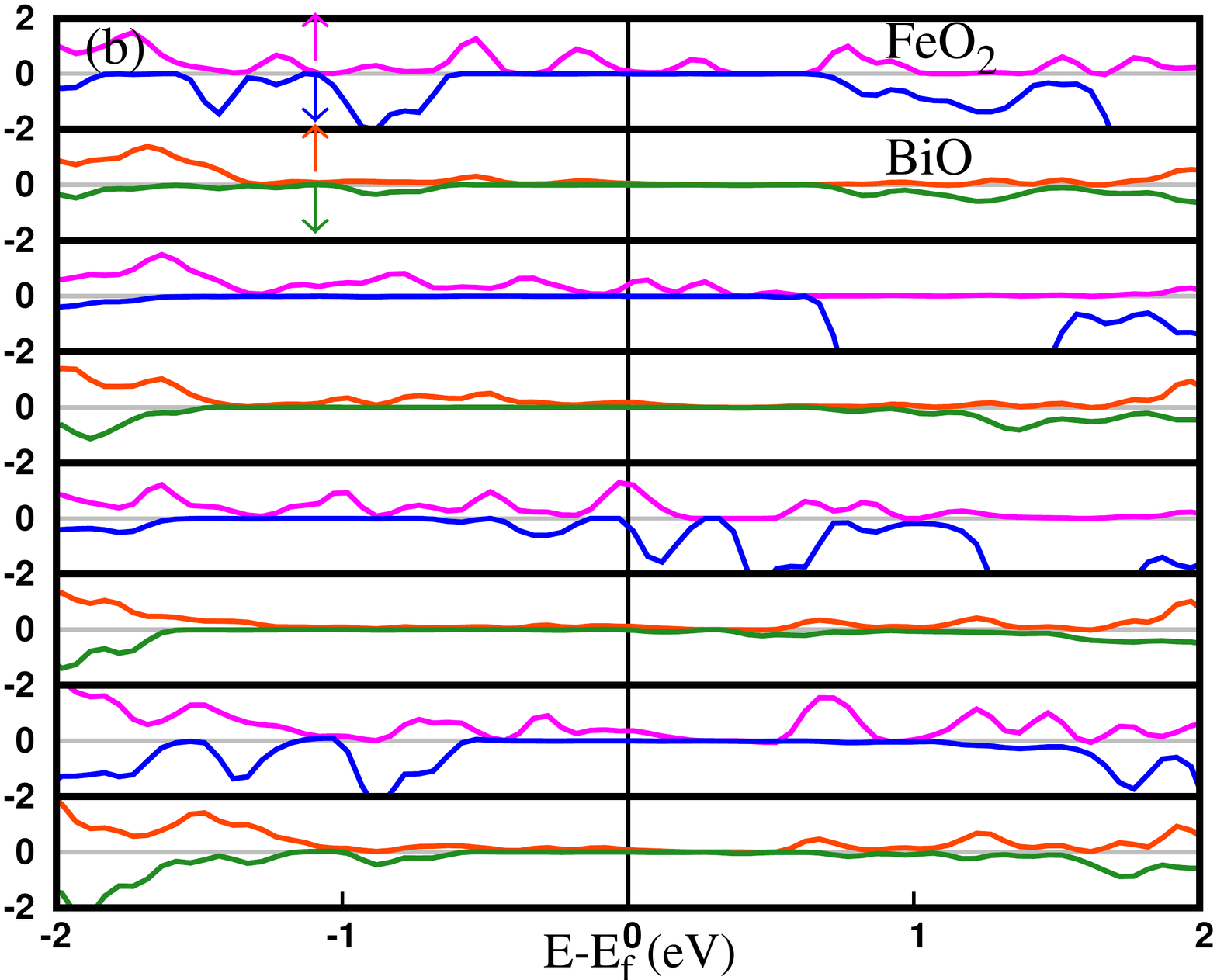}\hspace{-0.5cm}
	\includegraphics[width=0.265\textwidth,height=0.25\textwidth]{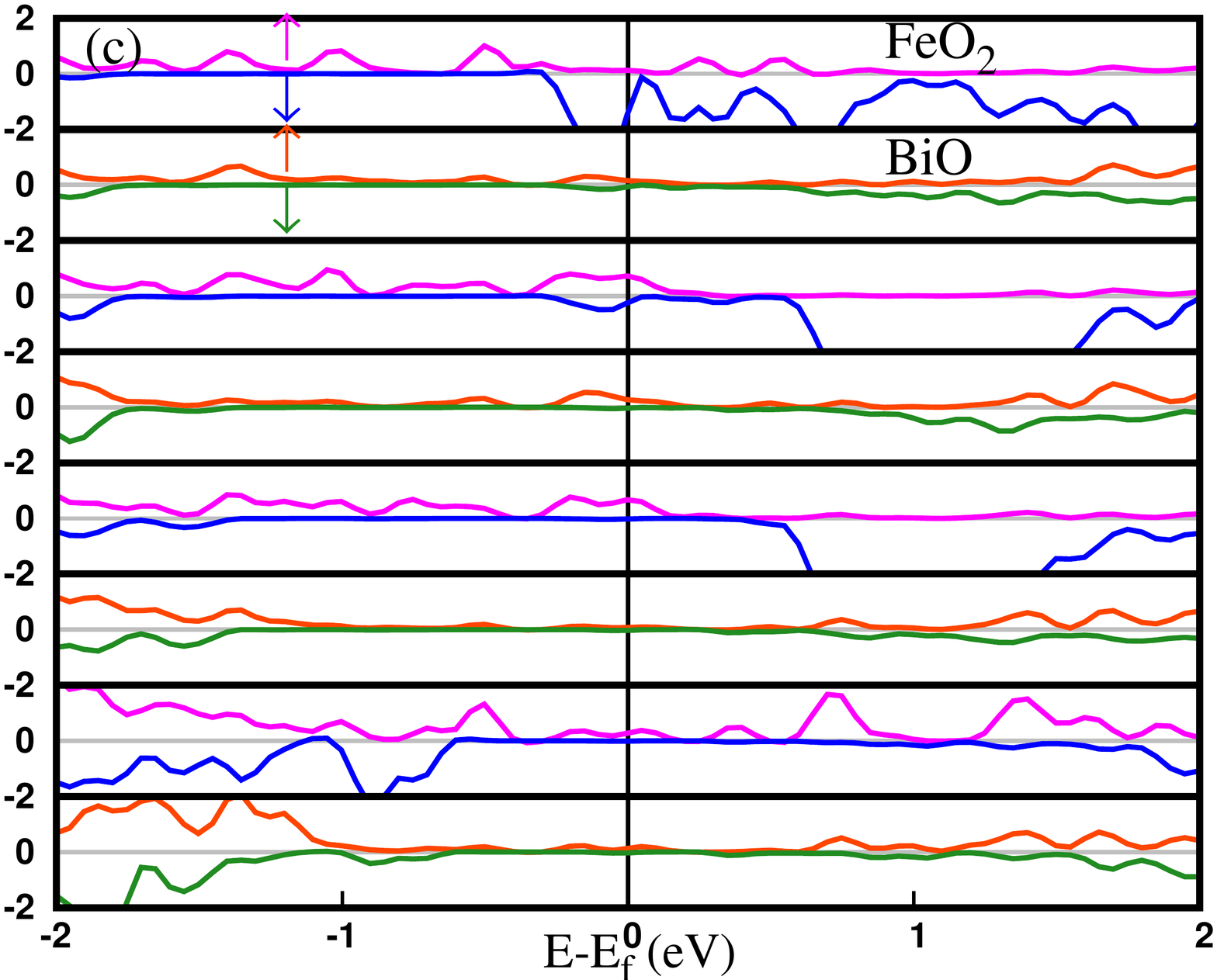}\hspace{-0.5cm}
	\includegraphics[width=0.265\textwidth,height=0.25\textwidth]{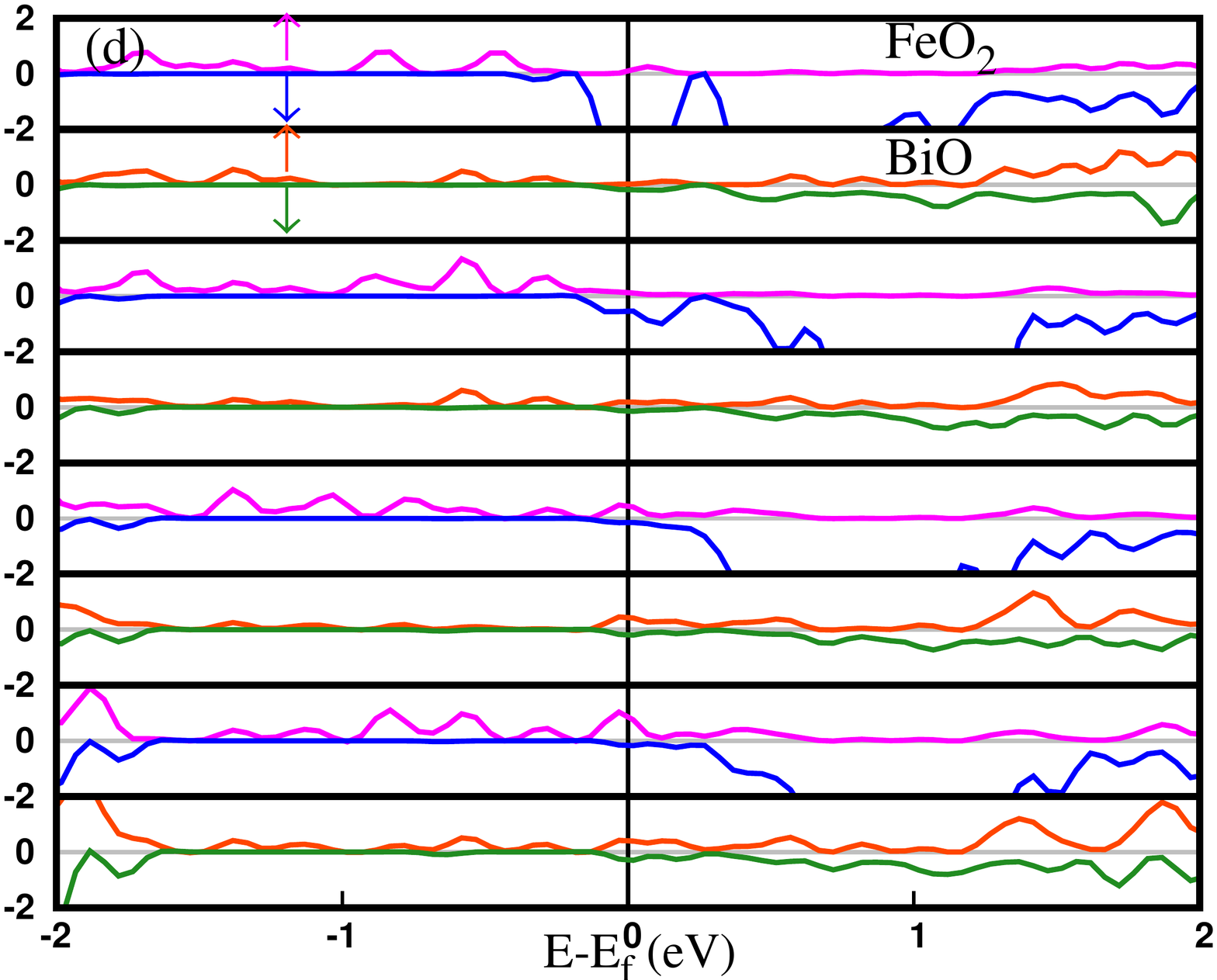}\hspace{-0.5cm}
\vspace{-0.4cm}
\caption{Density of states (DOS) of 4ml slab-thicknesses with $\rm{FeO_2}$ termination for (a) structure-I, (b) structure-II, 
(c) structure-III and (d) structure-IV respectively.} 
\label{Fig:fr-4l-dos-feo2}
\end{figure*}
\begin{figure*}[htbp!]
\centering
	\includegraphics[width=0.265\textwidth,height=0.25\textwidth]{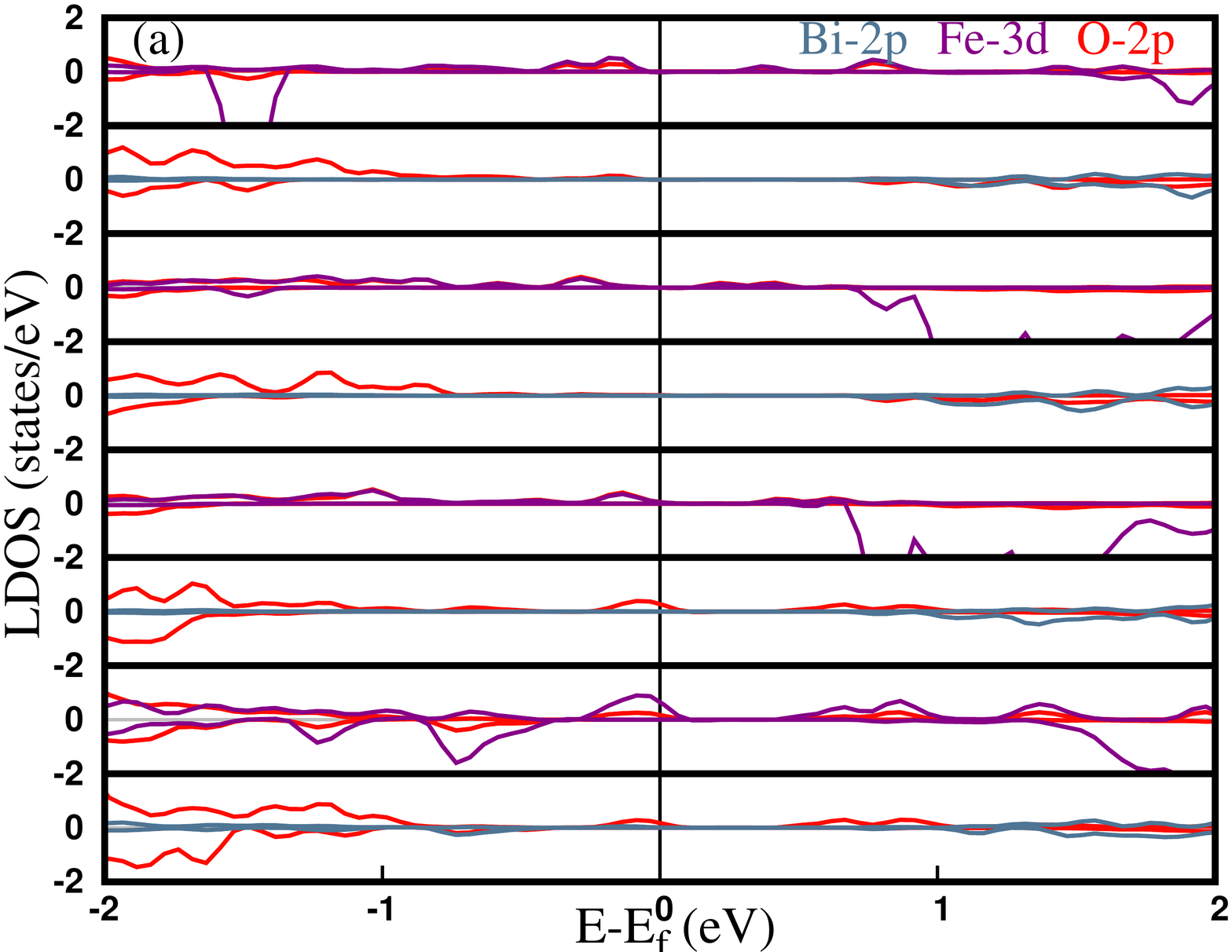}\hspace{-0.5cm}
	\includegraphics[width=0.265\textwidth,height=0.25\textwidth]{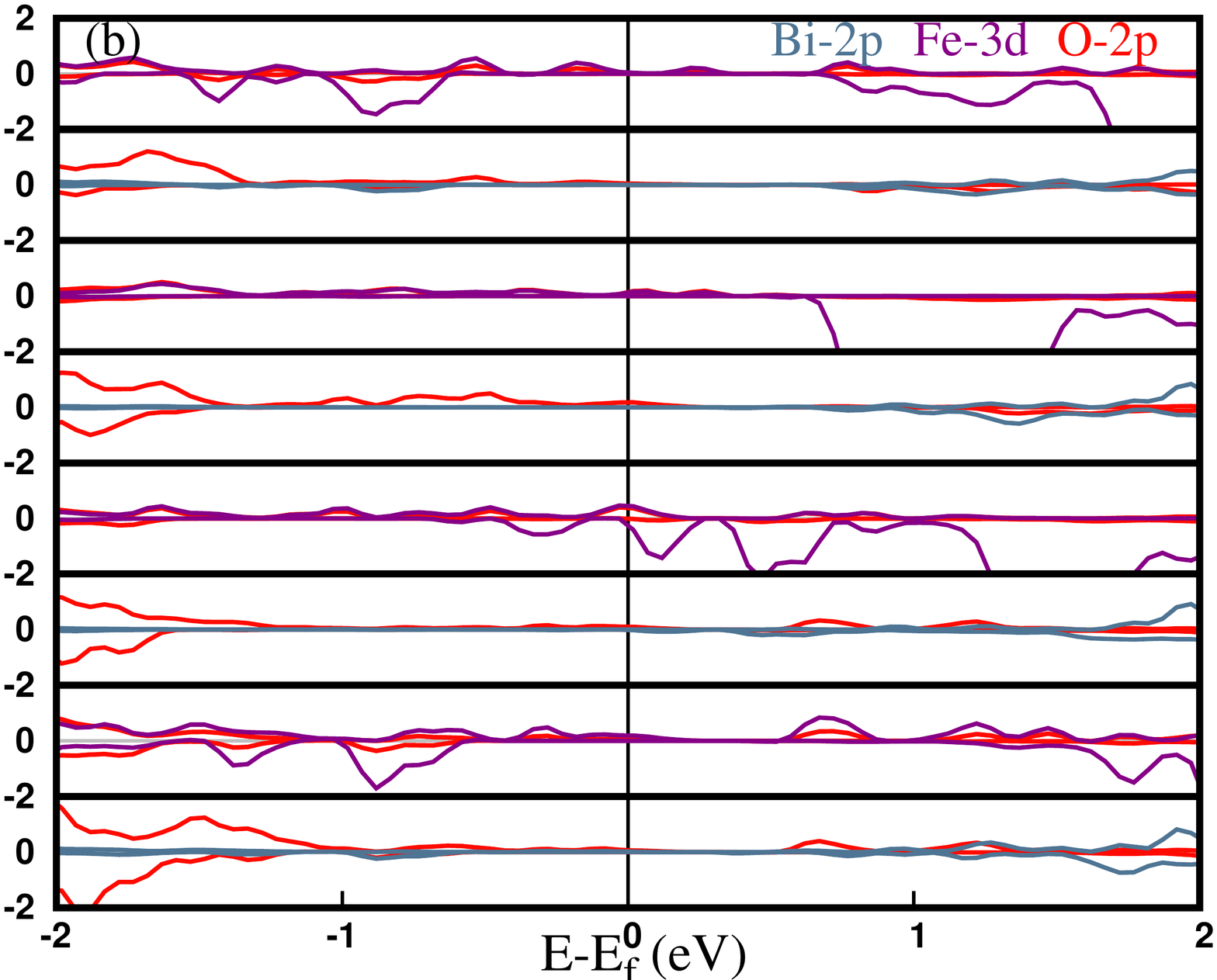}\hspace{-0.5cm}
	\includegraphics[width=0.265\textwidth,height=0.25\textwidth]{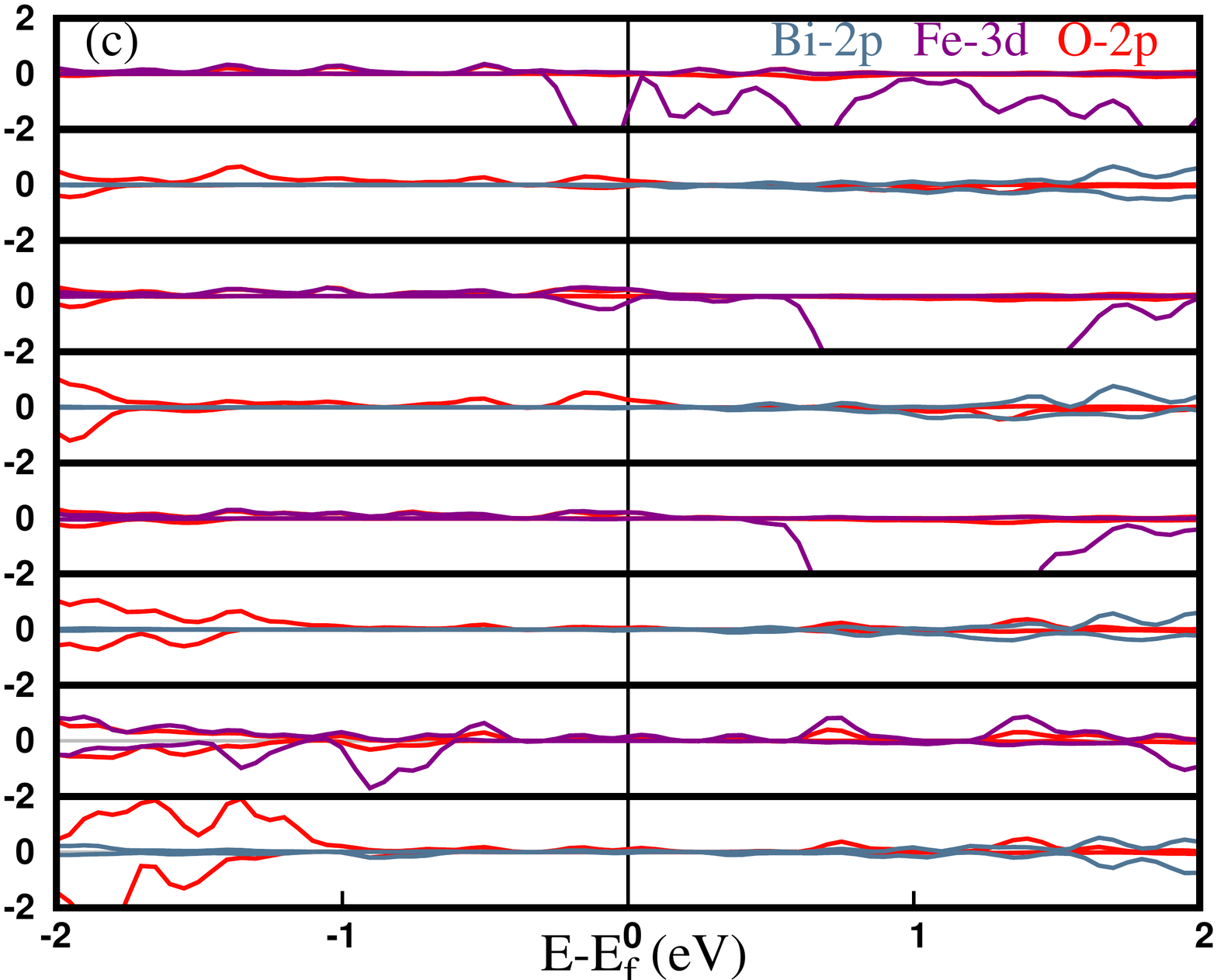}\hspace{-0.5cm}
	\includegraphics[width=0.265\textwidth,height=0.25\textwidth]{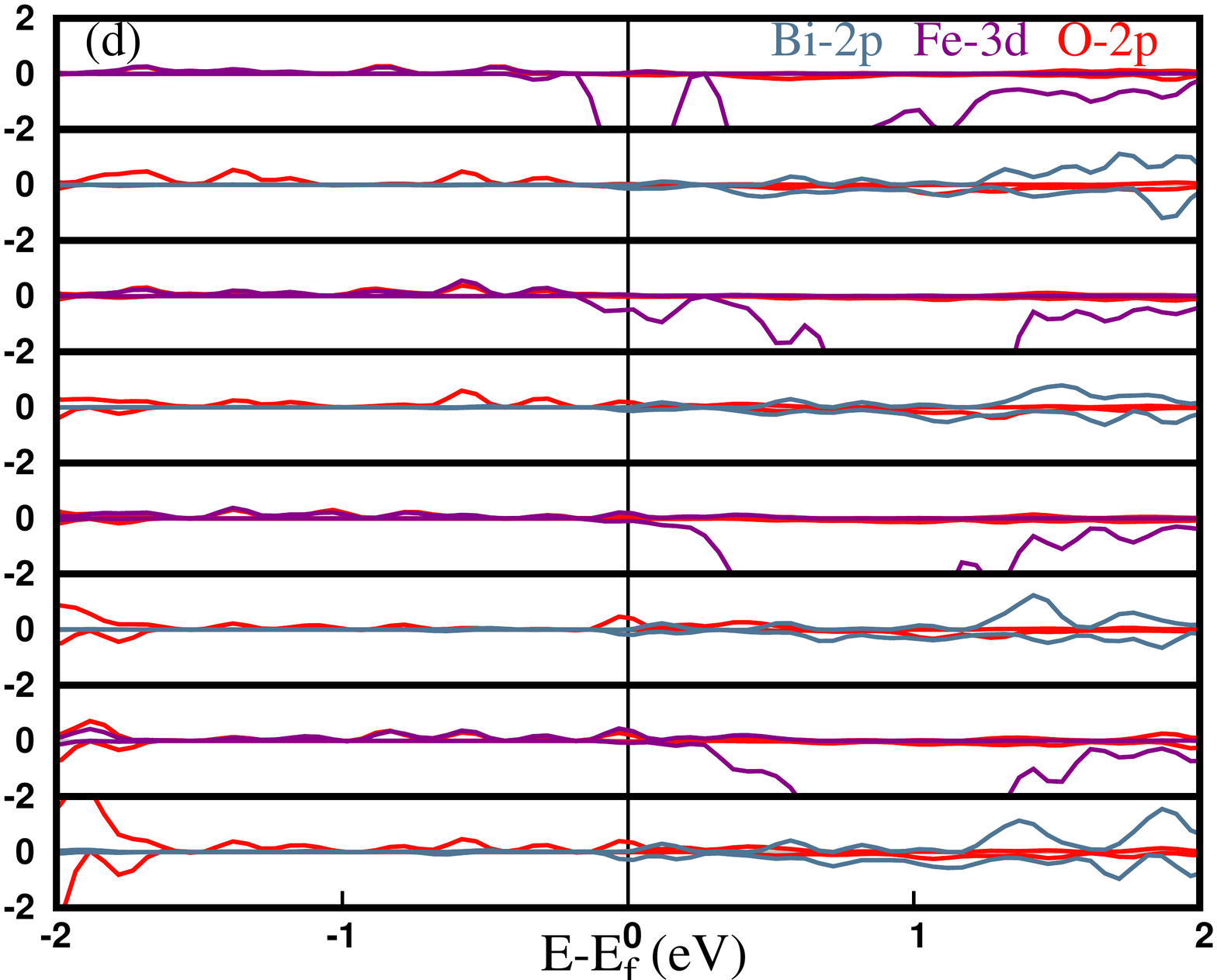}\hspace{-0.5cm}
\vspace{-0.4cm}
\caption{Local density of states(LDOS) of 4ml slab-thicknesses with $\rm{FeO_2}$ termination for (a) structure-I, (b) structure-II, 
(c) structure-III and (d) structure-IV respectively.} 
\label{Fig:fr-4l-ldos-feo2}
\end{figure*}
\section{Full-slab-study}\label{Appendix:full-slab}
In this section, the $\rm{FeO_2}$ terminated slabs with thicknesses 2ml and 4ml are presented. 
The results of DOS and LDOS results for full slab geometries where the whole slab
structures have been allowed to relax before investigating the electronic properties have been presented.
For 2ml thick slab, half-metallic electronic state and 2DHG have been found to persist in all the structures except structure-I. Surprisingly,
structure-I turns out to be metallic in this case. The layer resolved DOS for each structure have been presented in Fig.~\ref{Fig:fr-2l-dos-feo2}. 
From Fig.~\ref{Fig:fr-2l-dos-feo2}(a), it is evident that although the top surface of structure-I still 
exhibits the half-metallicity, the bottom layer shows metallic character. However, it is evident that for the 
\begin{table}[htbp!]
\caption{Estimation of the surface energies(in unit $\rm{J/m^2}$) for the $\rm{FeO_2}$ surface termination.}
\begin{center}
\begin{tabular}{c c c c c c c }
\hline
Structures  & str-I  & str-II & str-III & str-IV   \\
\hline
2ml          & 0.01 & 1.4  & 2.69 & 2.82\\
4ml         & -2.25 & 0.50  & 3.09 & 1.98 \\
\hline
\label{Table:feo2-relaxation-energy-fs}
\end{tabular}
\end{center}
\end{table}
rest of the structures  only one spin channel contributes at the Fermi energy. Furthermore, it is also 
clear that the charge carriers are hole type in theses structures. Surprisingly, however, for 4ml thick slab
the structure-I turns out to be a half-metal (Fig.~\ref{Fig:fr-4l-dos-feo2}(a)). However, from the thermodynamic
stability analysis (Table~\ref{Table:feo2-relaxation-energy-fs}) we find that the surface energy of this 
structure is negative, which indicates that this structure may not be stable. In case of structure-II, 
which is thermodynamically most stable a tiny contribution of spin-down electron carriers from the 
third $\rm{FeO_2}$ layer destroys the half-metallic character. In addition, structure-I, II still exhibit 
the existence of 2DHG. The 2DHG cease to exist in structure-III and IV. From Figs.~\ref{Fig:fr-4l-dos-feo2}(c) and (d), it 
is evident that the structure-III and IV behave as metal.
To understand the orbital contribution for each structure,  
we have also presented the LDOS results for the 2ml and 4ml thick slabs in Figs.~\ref{Fig:fr-2l-ldos-feo2} and \ref{Fig:fr-4l-ldos-feo2}. 
From the LDOS results we observe that Bi does not have contribution at the Fermi energy except for 2ml thick slab
of structure-I and 4ml thick structure-IV, which are metallic in nature, while contributions from $\rm{Fe-3d}$ and $\rm{O-2p}$
orbitals are found in both the half-metallic and mettalic  cases.
From the LDOS results (Fig.~\ref{Fig:fr-4l-ldos-feo2}(b) and (c)) we can observe that Fe-3d orbital contributes these electron carriers. 


\end{document}